\documentclass[english,pra,twocolumn,nopacs,preprintnumbers,eqsecnum,amsmath,amssymb,reprint]{revtex4}
\usepackage[T1]{fontenc}
\usepackage{verbatim}
\usepackage{amssymb}
\usepackage{graphicx}
\usepackage{times}
\usepackage{color}
\makeatletter

\newcommand{\dd}{\mathrm{d}}
\newcommand{\ee}{\mathrm{e}}
\newcommand{\rr}{\mathbf{r}}
\newcommand{\kk}{\mathbf{k}}

\@ifundefined{textcolor}{}
{%
 \definecolor{BLACK}{gray}{0}
 \definecolor{WHITE}{gray}{1}
 \definecolor{RED}{rgb}{1,0,0}
 \definecolor{GREEN}{rgb}{0,1,0}
 \definecolor{BLUE}{rgb}{0,0,1}
 \definecolor{CYAN}{cmyk}{1,0,0,0}
 \definecolor{MAGENTA}{cmyk}{0,1,0,0}
 \definecolor{YELLOW}{cmyk}{0,0,1,0}
}

\makeatother

\usepackage{babel}
\begin{document}
\let\vaccent=\v{
\global\long\def\gv#1{\ensuremath{\mbox{\boldmath\ensuremath{#1}}}}
\global\long\def\uv#1{\ensuremath{\mathbf{\hat{#1}}}}
\global\long\def\abs#1{\left| #1 \right|}
\global\long\def\avg#1{\left< #1 \right>}
\let\underdot=\d{
\global\long\def\pd#1#2{\frac{\partial#1}{\partial#2}}
\global\long\def\pdd#1#2{\frac{\partial^{2}#1}{\partial#2^{2}}}
\global\long\def\pdc#1#2#3{\left( \frac{\partial#1}{\partial#2}\right)_{#3}}
\global\long\def\op#1{\hat{\mathrm{#1}}}
\global\long\def\ket#1{\left| #1 \right>}
\global\long\def\bra#1{\left< #1 \right|}
\global\long\def\braket#1#2{\left< #1 \vphantom{#2}\right| \left. #2 \vphantom{#1}\right>}
\global\long\def\matrixel#1#2#3{\left< #1 \vphantom{#2#3}\right| #2 \left| #3 \vphantom{#1#2}\right>}
\global\long\def\av#1{\left\langle #1 \right\rangle }
 \global\long\def\com#1#2{\left[#1,#2\right]}
\global\long\def\acom#1#2{\left\{  #1,#2\right\}  }
\global\long\def\grad#1{\gv{\nabla} #1}
\let\divsymb=\div 
\global\long\def\div#1{\gv{\nabla} \cdot#1}
\global\long\def\curl#1{\gv{\nabla} \times#1}
\let\baraccent=\={

\title{Pseudo-thermalization in driven-dissipative non-Markovian open quantum systems}

\author{Jos\'e Lebreuilly}
\email{jose.lebreuilly@unitn.it}
\affiliation{INO-CNR BEC Center and Dipartimento di Fisica, Universit\`a di Trento, I-38123 Povo, Italy}

\author{Alessio Chiocchetta}
\affiliation{Institut f\"{u}r Theoretische Physik, Universit\"{a}t zu K\"{o}ln, D-50937 Cologne, Germany}

\author{Iacopo Carusotto}
\affiliation{INO-CNR BEC Center and Dipartimento di Fisica, Universit\`a di Trento, I-38123 Povo, Italy }

\begin{abstract}
We investigate a `pseudo thermalization' effect, where an open quantum system coupled to a non-equilibrated environment consisting of several non-Markovian reservoirs presents an emergent thermal behaviour. This thermal behaviour is visible at both static and dynamical levels and the system satisfies the fluctuation-dissipation theorem. Our analysis is focused on the exactly solvable model of a weakly interacting driven-dissipative Bose gas  in presence of frequency-dependent particle pumping and losses, and is based on a quantum Langevin theory, which we derive starting from a microscopical quantum optics model. For generic non-Markovian reservoirs, we demonstrate that the emergence of thermal properties occurs in the range of frequencies corresponding to low-energy excitations. For the specific case of non-Markovian baths verifying the Kennard-Stepanov relation, we show that pseudo-thermalization can instead occur at all energy scales. The possible implications regarding the interpretation of thermal laws in low temperature exciton-polariton and experiments are discussed.  We finally show that the presence of either a saturable pumping or a dispersive environment leads to a breakdown of the pseudo-thermalization effect.
\end{abstract}

\maketitle

\section{Introduction}

Our understanding of the conditions allowing for the emergence of equilibrium features in driven-dissipative quantum systems is still incomplete. The dynamics of open quantum systems is often characterized by the presence of a complex external environment, implementing a wide range of effects such as single particle and many-body losses, pump, dephasing \cite{Carusotto_rev,Hartmann_rev}, or more exotic dissipative processes \cite{Diehl_atom_dissipation}, which are usually modelled as a series of external reservoirs \cite{Feynman_influence,Caldeira_dissipation}. Due to the presence of dissipation, in the generic situation an open quantum system is expected to reach after a long enough evolution a steady-state where observables no longer evolve in time \cite{Breuer,Quantum_Noise}. Although it is a widely accepted belief that in presence of a typical non-equilibrated environment the system properties do not necessarily recover those predicted by some thermal model, a quantitative  estimation of the deviations between the steady-state and equilibrium predictions often reveals challenging.

Over the last decade, these problematics have become particularly relevant at an experimental level also in the quantum regime, as pioneering works in photonic devices have opened a whole new research direction on the dynamics of non-equilibrium quantum fluids. Signatures of Bose-Einstein distributions, such as the presence of power-law infrared divergencies similar to the Rayleigh-Jeans distribution ($n_k\propto k^{-2}$ for $k\to0$), and/or high-energy exponential tails of a Boltzmann type ($n_k\propto\text{exp}[-\beta E_k]$ for $k\to\infty$), have been observed in several experiments involving photon and exciton-polariton non-equilibrium gases \cite{Kasprzak_BEC,Balili_BEC,Bajoni_BEC,Klaers_thermalization,Klaers_BEC_dye,Kena_room_BEC,Plumhof_room_BEC,Schmitt_thermalization_kinetics}. In room temperature experiments \cite{Klaers_thermalization,Klaers_BEC_dye,Kena_room_BEC,Plumhof_room_BEC,Schmitt_thermalization_kinetics}, the appearance of  thermal correlations might be seen as something rather predictible since energy exchange with the thermal environment is occurring much faster than particle losses. Yet, in other classes of low-temperature exciton-polaritons \cite{Kasprzak_BEC,Balili_BEC} and VCSEL \cite{Bajoni_BEC} experiments where non-equilibrium effects are expected to be kinetically dominant, the underlying mechanisms leading to the emergence of an effective temperature differing from the one of the apparatus are less clear and subject to controversy \cite{Porras_scattering,Malpuech_exciton_scattering,Maragkou_optical_phonons}.

From a theoretical point of view, many studies have quantified the distance from equilibrium for photonic systems \cite{Porras_scattering,Carusotto_LR_order,Chiocchetta_Wigner,Chiocchetta_Langevin,Kirton_BEC,Kirton_breakdown_thermalization}. In \cite{DellaTorre} it was shown that the presence of suitably designed $1/f$ noise in a generic open quantum system could lead to critical properties analogous to an equilibrium quantum phase transition. Works based on the renormalization group (RG) \cite{DellaTorre_RG,Sieberer_3D,Sieberer_BEC} and diagrammatic expansions \cite{DellaTorre_diagrammatic} for non-equilibrium field theories have addressed the long-range and low-energy properties of quantum fluids and the critical properties across a driven-dissipative phase transition, and connections have been drawn between equilibrium and symmetries of the Keldysh action \cite{Aron_Symmetries,Sieberer_Symmetries}.  In particular, the important role played by the spatial dimensionality in determining whether a driven-dissipative quantum system presents asymptotic thermal properties was pointed out in many studies \cite{Sieberer_3D,Altman_2D,Sieberer_2D_vortices,He_1D}. More recently, the necessity of characterizing the dynamical properties was highlighted in \cite{Chiocchetta_thermalization}, where it was showed that a driven-dissipative quantum system could present at steady-state equilibrium-like static correlations without verifying the fluctuation-dissipation theorem (FDT)  \cite{Kubo_FDT} at a dynamical level.

Here we want to push this last statement one step further: we argue that, under specific conditions, an open quantum system can present all the attributes of an equilibrated system both at a static and a dynamic level, verifying thus the FDT theorem, even though its environment is highly non-thermal. In a previous work \cite{Lebreuilly_2016}, we unveiled a preliminary result in this direction for a quantum optical model, where we showed that apparent thermalization can be obtained by coupling the system to several non-thermal and non-Markovian baths, which effectively mimic the impact of a single thermal bath. 

Thermal signatures have already been predicted to emerge in high enough dimensions in the long-range behaviour of generic interacting non-equilibrium systems  \cite{Sieberer_3D,Sieberer_BEC} and in Rydberg atoms  in presence of a suitably engineered environment \cite{Schonleber_artificial}. Furthermore, hints toward the validity of some fluctuation-dissipation relations in driven-dissipative quantum spin systems were recently found in \cite{Keeling_thermalization}. Beyond these works, we stress that the effective equilibrium predicted here relies on a different physical mechanism: whenever the Kennard-Stepanov (KS) relation \cite{Kennard,Stepanov} (i.e., a particular form of detailed balance relation) is verified in our model, the system is not able to perceive that the reservoirs are not equilibrated and its steady-state coincides with a thermal state, with both temperature and chemical potential being emergent quantities depending on the spectral properties of the various baths. We choose to call this effect \textit{``pseudo-thermalization''}.

Following \cite{Lebreuilly_2016}, the preliminary concept was deepened in \cite{Shabani_artificial_thermal_bath}, who suggested to engineer more complex reservoirs so to reproduce this mechanism over broader energy scales, and then obtain artificial and controllable temperatures in view of optimizing the performance of quantum annealers. Some hints suggest that the apparent emergence of thermal static properties in low-T exciton-polariton \cite{Kasprzak_BEC,Balili_BEC} and VCSEL \cite{Bajoni_BEC} experiments  might be related to pseudo-thermalization in some experimental configurations. In the very recent work \cite{Lebreuilly_square}, two of us suggested to exploit a closely related effect to stabilize photonic Mott Insulating states close to zero temperature.

In both works \cite{Lebreuilly_2016,Shabani_artificial_thermal_bath}, the formalism was based on a quantum master equation formalism, which allowed to compute the static properties of the steady-state. However, due to the absence of a regression theorem for non-Markovian problems \cite{Breuer,Guarnieri_regression_non-markov}, such approach does not allow to access dynamical physical quantities such as multiple time correlators, and in particular is not suited to verify the validity of the fluctuation-dissipation theorem. Moreover, as all predictions were based on very general theoretical arguments, a full validation on an exactly solvable model still remains to provide. 

In this paper, we investigate pseudo-thermalization effects for the specific model of a weakly interacting BEC coupled to several non-Markovian reservoirs. In contrast with \cite{Lebreuilly_2016,Shabani_artificial_thermal_bath} we develop an alternative analytical approach based on a quantum Langevin formalism which keeps tracks of the bath dynamics and in particular allows to access both static and dynamical properties of the steady state. In this way, we are able not only to demonstrate the presence of thermal signatures at a static level, but also to show that the fluctuation-dissipation theorem is verified at a dynamical level.

This paper is organized as follows: in Sec.~\ref{sec:quantum-Langevin} we introduce the general Langevin model and use a Bogoliubov approach to linearize the theory around a mean-field solution, from which we demonstrate numerically the dynamical stability. We also derive a low-energy  effective description, allowing to provide exact analytical expressions for the low-momentum Bogoliubov spectrum. In Sec.~\ref{sec:effective-thermalization}, we show that, for baths with arbitrary spectral shape, this model presents low-energy pseudo-thermalization both at a static and dynamical level: we demonstrate that at low energies not only static correlations match with their thermal counterpart, but that the FDT is also verified. Moreover, if the non-thermal baths are suitably chosen to verify the Kennard-Stepanov (KS) relation at all energies, then the system undergoes thermalization at all energies. In Sec.~\ref{sec:Langevin-derivation} we provide a microscopic derivation of the quantum Langevin model starting from a quantum optical model involving frequency-dependent losses and emitters with a non-trivial distribution of transition frequencies. We also explain how the Kennard-Stepanov relation could be engineered with this model, and how it might be naturally reproduced in some specific low-T exciton-polariton  and VCSEL experiments. In Sec.\ref{sec:out-of-equilibrium} we give hints on how pseudo-thermalization can be broken and the system be driven out-of-equilibrium by adding saturation and/or non-trivial momentum dependence to the dissipative processes responsible for particle pumping. Conclusions are given in Sec.~\ref{sec:conclusions}.
\section{Non-Markovian quantum-Langevin equation}
\label{sec:quantum-Langevin}
In this section we introduce a theoretical model for the dynamics of a driven-dissipative interacting Bose Gas in contact with non-Markovian reservoirs. A similar model had already been addressed in a quantum optics context in \cite{Lebreuilly_2016,Lebreuilly_square}  but it was formulated in terms of a Redfield master equation instead of the quantum Langevin formalism used here. Focusing on the weakly interacting case, in the BEC regime we study the mean-field solution of this model and use the Bogoliubov theory to study the dynamics of fluctuations. After demonstrating numerically the dynamical stability for a specific choice of the pump and loss spectra, we develop a low-energy effective theory so to access analytically the low-momentum collective modes of the condensate. 
\subsection{Model for a driven condensate}
\label{sec:model}
Let us consider a bosonic gas in $d$ spatial dimensions, described by the annihilation and creation fields $\hat{\psi}(\rr)$ and $\hat{\psi}^\dagger(\rr)$. The evolution in time of these operators is described by the non-Markovian quantum-Langevin equation
\begin{multline}
\label{eq:quantum-langevin-Bose}
\frac{\partial \hat{\psi}}{\partial t}(\rr,t) = -i\left[\omega_0 - \frac{\nabla^2}{2m}+g \hat{\psi}^\dagger(\rr,t)\hat{\psi}(\rr,t) \right]\hat{\psi}(\rr,t)\\+ \int_{t'} \Gamma(t') \hat{\psi}(\rr,t-t') 
+ \hat{\xi}(\rr,t),
\end{multline} 
where $\int_t' \equiv \int_{-\infty}^{+\infty}\dd t'$, while $\omega_0$ is the bare cavity frequency, $m$ is the bosonic mass, $g>0$ is the strength of the repulsive contact interaction, $\Gamma$ is a memory kernel and $\hat{\xi}(\rr,t)$ a zero-mean Gaussian quantum noise operator. Equation~(\ref{eq:quantum-langevin-Bose}) resembles the Heisenberg equation for the motion of the  operator $\hat{\psi}$ for an isolated interacting Bose gas. However, the dynamics described by Eq.~(\ref{eq:quantum-langevin-Bose})  does not conserve energy and number of particles. Namely,  the memory kernel $\Gamma(t')$ and quantum noise $\hat{\xi}(t)$ terms model altogether the effect of non-Markovian particle loss  and incoherent pumping (i.e., injection) processes, whose respective strength is quantified by the frequency-dependent power spectra $\mathcal{S}_{\rm{l}}(\omega)$ and $\mathcal{S}_{\rm{p}}(\omega)$. 

Within the Langevin formalism, the correlations of the noise operators $\hat{\psi}(\rr,t)$, $\hat{\psi}^\dagger(\rr,t)$ can be written as  
\begin{subequations}
\label{eq:noise-correlation}
\begin{align}
\langle \hat{\xi}(t)\hat{\xi}^{\dagger}(t')\rangle & =  \int_\omega \mathcal{S}_{\rm{l}}(\omega)\,\ee^{-i\omega (t-t')}\label{eq:noise-correlation-loss} \\
\langle \hat{\xi}^{\dagger}(t)\hat{\xi}(t')\rangle & =  \int_\omega \mathcal{S}_{\rm{p}}(\omega)\, \ee^{i\omega (t-t')} \label{eq:noise-correlation-pump},
\end{align}
\end{subequations}
with $\int_{\omega}\equiv \int_{-\infty}^{+\infty} \dd\omega/(2\pi)$. Likewise, $\Gamma$ is expressed as
\begin{equation}
\label{eq:memory-kernel}
\Gamma(t) = \theta(t)\int_\omega \left[\mathcal{S}_{\rm{p}}(\omega) -\mathcal{S}_{\rm{l}}(\omega)  \right]\ee^{-i\omega t}.
\end{equation}
The Heaviside function $\theta(t)$ in Eq.~(\ref{eq:memory-kernel}) is needed in order to ensure causality: as a result, its presence implies the Kramers-Kronig relations between the real and imaginary parts of the Fourier transform $\Gamma(\omega) = \int_t \ee^{i \omega t} \Gamma(t)$, which can thus be written as
\begin{subequations}
\label{eq:KK-relations}
\begin{align}
\text{Re}\left[\Gamma(\omega)\right] &=\frac{1}{2}\left[\mathcal{S}_{\rm{p}}(\omega)-\mathcal{S}_{\rm{l}}(\omega)\right], \\
\text{Im}\left[\Gamma(\omega)\right] &=\text{PV}\int_{\omega'} \frac{\mathcal{S}_{\rm{p}}(\omega')-\mathcal{S}_{\rm{l}}(\omega')}{\omega-\omega'}. 
\end{align}
\end{subequations}

The power spectra $\mathcal{S}_{\rm{p}}(\omega)$ and $\mathcal{S}_{\rm{l}}(\omega)$ are assumed to be smooth functions of the frequency $\omega$. In the following, we will restrict to the case in which there exists a range of frequencies $\omega_1 < \omega <\omega_2$ such that $\mathcal{S}_{\rm{p}}(\omega) > \mathcal{S}_{\rm{l}}(\omega)$ (``amplifying'' region), and that $\mathcal{S}_{\rm{p}}(\omega) < \mathcal{S}_{\rm{l}}(\omega)$ outside this interval (``lossy'' region). Accordingly, losses are perfectly balanced by pumping at the boundary of this interval, i.e.,  $\mathcal{S}_{\rm{p}}(\omega_{1,2}) = \mathcal{S}_{\rm{l}}(\omega_{1,2})$. 
We also define 
\begin{equation}
\Delta_{\rm{diss}}=\text{min}(\text{FHWM}(\mathcal{S}_{\rm{l}}),\text{FHWM}(\mathcal{S}_{\rm{p}})) \label{eq:delta_diss}
\end{equation}
as the minimum of the full width at half maximum of the power spectra $\mathcal{S}_{\rm{l}}(\omega) $ and $\mathcal{S}_{\rm{p}}(\omega) $. It represents a characteristic frequency scale over which these power spectra change value and quantifies the non-Markovianity of the dynamics. 

We stress that the loss and pump power spectra $\mathcal{S}_{\rm{l}}(\omega)$ and $\mathcal{S}_{\rm{p}}(\omega)$ arise from the contact of the system with separate reservoirs, i.e., a lossy medium and an amplifying medium (these reservoirs are respectively composed of pure absorbers and pure emitters): as a consequence, $\mathcal{S}_{\rm{l}}(\omega)$ and $\mathcal{S}_{\rm{p}}(\omega)$ are assumed to be perfectly independent and completely tunable physical quantities. A microscopic derivation based on a quantum optical model of the quantum Langevin Equation (\ref{eq:quantum-langevin-Bose}) illustrating all these features  is presented in Sec.~\ref{sec:Langevin-derivation}.

Finally we introduce the following quantity 
\begin{equation}
\label{eq:effective-temperature}
\beta_\text{eff}\equiv\frac{1}{T_{\text{eff}}}
\equiv\frac{S'_{\rm{l}}(\omega_2)-S'_{\rm{p}}(\omega_2)}{\mathcal{S}_{\rm{p}}(\omega_2)}  = \frac{\dd }{\dd \omega}\log\left[\frac{\mathcal{S}_{\rm{l}}(\omega)}{\mathcal{S}_{\rm{p}}(\omega)}\right]\biggr|_{\omega = \omega_2}.
\end{equation}
As we will see in Sec.~\ref{sec:effective-thermalization}, this model presents pseudo-thermalization properties at low energies for generic power spectra, and $T_{\text{eff}}$ will play the role of an effective temperature. $T_{\text{eff}}$ also scales like the linewidth  $\Delta_{\rm{diss}}$ of the power spectra defined in Eq.~(\ref{eq:delta_diss})  and quantifies the non-Markovianity of the dissipative dynamics, but unlike  $\Delta_{\rm{diss}}$ it is more sensitive to the local properties in frequency space around $\omega_2$. In the Markovian limit, the power spectra are very flat and we have that $T_{\text{eff}},\Delta_{\rm{diss}}\to\infty$. On the contrary, for very steep power spectra  (very coherent pump and/or loss processes), the dynamics is highly non-Markovian and we have that $T_{\text{eff}},\Delta_{\rm{diss}}\to 0$. 

In analogy with what was already discussed in \cite{Lebreuilly_2016,Shabani_artificial_thermal_bath}, here the physical origins of the pseudo-thermalization can be understood intuitively at a qualitative level: at  $\omega_2$ losses and pump exactly compensate ($\mathcal{S}_{\rm{p}}(\omega_2)= \mathcal{S}_{\rm{l}}(\omega_2)$), so  this frequency will play the role of the condensate frequency for this model. As we shall see below, a condensate at $\omega_1$ would instead be unstable. Modes at frequencies close to $\omega_2$ will correspond to low-energy excitations on top of the condensate. 

In the vicinity of $\omega_2$, the pump and loss power spectra verify the following condition (see Eq.~(\ref{eq:effective-temperature})):
\begin{equation}
\label{eq:approximate-KS}
\frac{ \mathcal{S}_{\rm{p}}(\omega_2+\omega)}{\mathcal{S}_{\rm{l}}(\omega_2+\omega)}\underset{\omega\to 0}{\simeq}(1-\beta_{\text{eff}}\omega+\mathcal{O}(\omega/\Delta_{\text{diss}})^2)\sim e^{-\beta_{\text{eff}}\omega},
\end{equation} 
so the Kennard-Stepanov relation \cite{Kennard,Stepanov} is asymptotically verified at low frequencies. Thus, as we will demonstrate in Sec.~\ref{sec:effective-thermalization}, steady-state low-energy properties are expected to be thermal.

Moreover, if we choose the pump and loss spectra to verify exactly the Kennard-Stepanov relation
\begin{equation}
\label{eq:exact-KS}
\frac{ \mathcal{S}_{\rm{p}}(\omega_2+\omega)}{\mathcal{S}_{\rm{l}}(\omega_2+\omega)}=e^{-\beta_{\text{eff}}\omega},
\end{equation} 
then the system should thermalize at all energies. Note that this can be obtained without the various reservoirs being at thermal equilibrium, as we can tune independently the  power spectra $S_{(\rm{l}/\rm{p})}$ by changing the frequency distributions of the excitations within the reservoirs respectively responsible for particle losses and pumping. In Sec.~\ref{sec:Langevin-derivation} we  will discuss a few physical contexts where the Kennard-Stepanov may be fulfilled.}

While obtaining a full thermalization requires a fine tuning of the reservoirs power spectra in order to fully verify the KS relation, all the results presented in the next sections regarding low-energy properties are general in the sense that they do not depend on the precise shape of the power spectra. In order to make our discussion concrete, we performed numerical simulations for a specific choice of $\mathcal{S}_{(\rm{l}/\rm{p})}(\omega)$. For all the graphical representations we will thus consider the case of Markovian losses and a Lorentzian-shaped pump (see Fig.~\ref{fig:power-spectra})
\begin{subequations}
\label{eq:lorentzian-Markovian-power-spectra}
\begin{align}
\mathcal{S}_{\rm{l}}^{graph}(\omega)&\equiv\Gamma_{l},\\
\mathcal{S}_{\rm{p}}^{graph}(\omega)&\equiv\Gamma_{\rm{p}}\frac{(\Delta_{\text{diss}}/2)^2}{(\omega-\omega_{\rm{p}})^2+(\Delta_{\text{diss}}/2)^2}.
\end{align} 
\end{subequations}
where the use of the notation $\Delta_{\text{diss}}$ is consistent with the previous definition. We also define the detuning $\delta\equiv\omega_0-\omega_{\rm{p}}$ between the photonic and the pump frequency. Accordingly, we need to have $\Gamma_{\rm{l}}<\Gamma_{\rm{p}}$ in order to obtain an amplified range of frequencies and generate a condensate, and $\omega_{1,2}$ are the two solutions of 
\begin{equation}
\frac{(\Delta_{\text{diss}}/2)^2}{(\omega-\omega_{\rm{p}})^2+(\Delta_{\text{diss}}/2)^2}=\frac{\Gamma_{\rm{l}}}{\Gamma_{\rm{p}}}.\end{equation} 
This choice of loss and pump power spectrum is naturally reproduced by our quantum optics proposal Sec.~\ref{subsec:engineer-lorentzian-spectrum}. Since it does not verify exactly the Kennard-Stepanov relation, we do not expect it will lead to complete thermalization; however, it is well suited to investigate the effect of low-energy pseudo-thermalization.\\

\begin{figure}[t]
\centering
\includegraphics[width=1\columnwidth,clip]{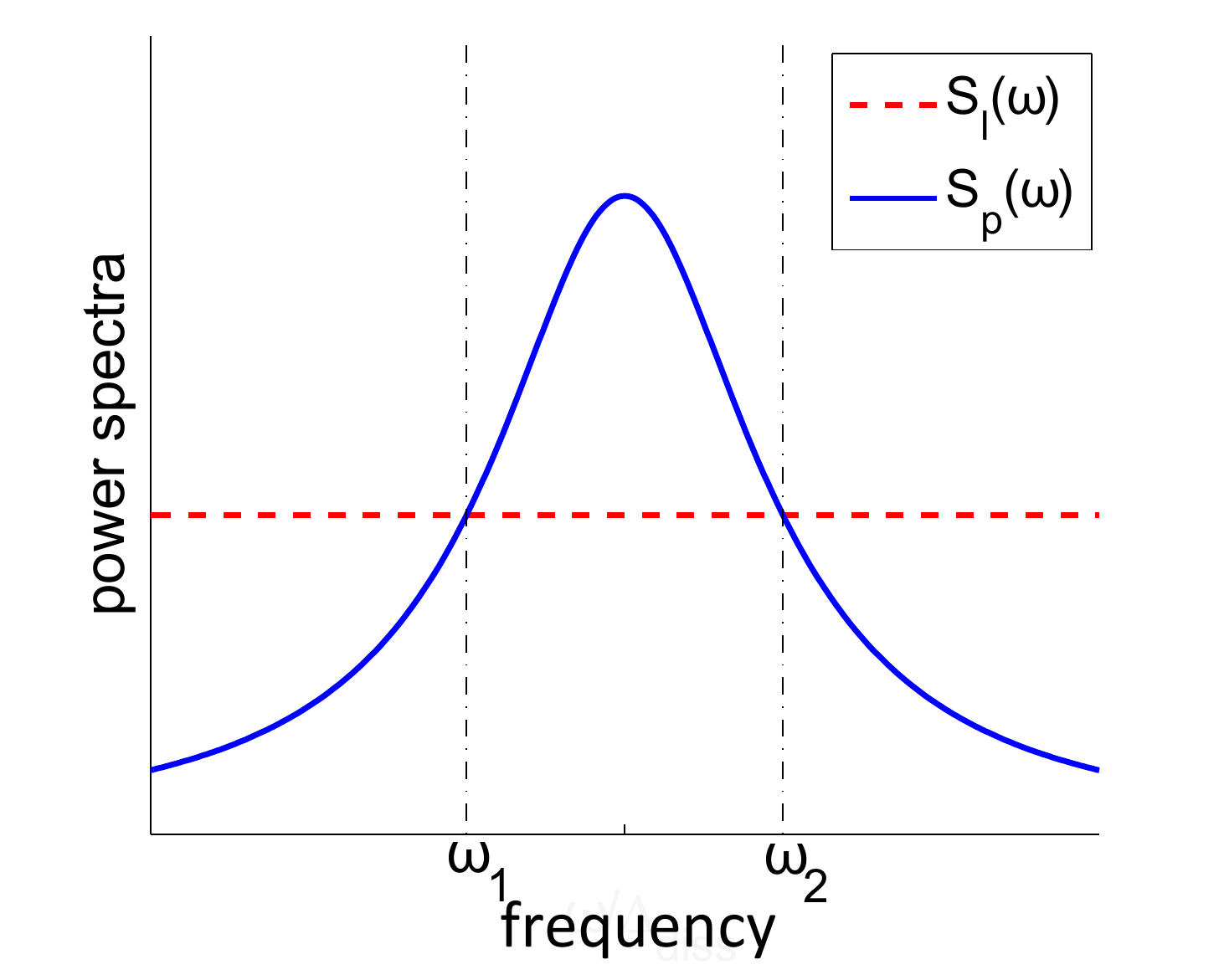}
\caption{\label{fig:power-spectra}Power spectra for Markovian losses and Lorentzian shape pump in arbitrary units}
\end{figure}
\subsection{Non-interacting case}
\label{sec:free-bose}
In this section we consider the case of a non-interacting Bose gas, i.e, we set the interaction strength $g=0$. 
In this case, the Langevin equation Eq.~(\ref{eq:quantum-langevin-Bose}) is linear and it can be solved exactly, for a given choice of $\Gamma(\omega)$. 
If a stationary state exists independent on the initial conditions (see discussion further below), one may evaluate the corresponding solution by introducing the Fourier transforms
\begin{subequations}
\label{eq:Fourier-definitions}
\begin{align}
\hat{\psi}_\kk(\omega) & = \int_{\rr, t} \hat{\psi}(\rr,t)\ee^{i(\kk\cdot \rr-\omega t)},\\
\hat{\psi}^\dagger_\kk(\omega) & = \int_{\rr, t} \hat{\psi}^\dagger(\rr,t)\ee^{-i(\kk\cdot \rr-\omega t)} =\left[\hat{\psi}_\kk(\omega) \right]^\dagger,\\
\hat{\xi}_\kk(\omega)& = \int_{\rr, t} \hat{\xi}(\rr,t)\ee^{i(\kk\cdot \rr-\omega t)},\\
\hat{\xi}^\dagger_\kk(\omega) & = \int_{\rr, t} \tilde{\xi}^\dagger(\rr,t)\ee^{-i(\kk\cdot \rr-\omega t)} = \left[\hat{\xi}_\kk(\omega) \right]^\dagger,
\end{align}
\end{subequations}
and by replacing them into Eq.~(\ref{eq:quantum-langevin-Bose}): one thus finds that the value of $\hat{\psi}_\kk(\omega)$ is given by
\begin{equation}
\label{eq:non-interacting-solution}
\hat{\psi}_\kk(\omega) = \frac{i\hat{\xi}_\kk(\omega)}{\omega-\omega_0 -\epsilon_\kk - i \Gamma(\omega)},
\end{equation}
with $\epsilon_\kk = k^2/2m$. Note that, as a consequence of the absence of the non-linearity, all the modes $\kk$ are decoupled.
When $\hat{\psi}_\kk(\omega)$ is transformed back in real time, it results in a linear combination of several modes $\omega_{\kk,n}$, corresponding to the poles of the denominator in Eq.~(\ref{eq:non-interacting-solution}), weighted with different amplitudes. For each value of $\kk$, several solutions $\omega_{\kk,n}$ (labelled by the index $n$) may exist: this give rise to a branched spectrum of eigenfrequencies. The number of these branches depends on the peculiar choice of $\Gamma(\omega)$: these additional branches account for the existence of external reservoir degrees of freedom which were integrated out in order to provide the dynamical description Eq.~(\ref{eq:quantum-langevin-Bose}) of the bosonic field $\hat{\psi}$.

The imaginary part $\text{Im}[\omega_{\kk,n}]$ corresponds to the inverse lifetime of the given mode: in order to have a dynamically stable mode, the condition $\text{Im}[\omega_{\kk,n}]<0$ must be satisfied; this also implies that a dynamically stable stationary solution independent of the initial state exists, as any information on the initial state will vanish exponentially fast in time. On the contrary, if $\text{Im}[\omega_{\kk,n}] \geq 0$ for some values of $\kk$ and $n$, the corresponding mode grows indefinitely in time, or it remains constant: in both cases, one cannot neglect the information about the initial state, thus invalidating the assumption that a stationary value independent on the initial state exist. For $\text{Im}[\omega_{\kk,n}] > 0$, the field $\hat{\psi}$ diverges exponentially in time, and thus the solution is physically meaningless: nonetheless, this feature may signal a dynamical instability of the non-interacting approximation of Eq.~(\ref{eq:quantum-langevin-Bose}), and, as a result, the inclusion of non-linearity may be crucial.   

For the choice of the power spectra discussed in Sec.~\ref{sec:model}, which admits an amplifying region $[\omega_1,\omega_2]$, one expects some eigenmodes to present dynamical instabilities. Qualitatively, if $\omega_0 +\epsilon_\kk$ falls into the amplifying region (which can be shifted with respect to $[\omega_1,\omega_2]$, due to the presence of the imaginary part $\text{Im}[\Gamma(\omega)]$ which induces a Lamb shift of the bare frequency), a dynamical instability is expected: while in a standard laser the instability would be controlled and ultimately stopped due to the presence of a saturated gain medium \cite{Scully_book,Mandel_book}, here those nonlinear terms where not included in our Langevin description. We will see below that the inclusion of a non-vanishing interaction strength $g\neq 0$ provides a non-standard saturation mechanism which prevents the unconstrained growth of dynamically unstable modes. 

\subsection{Interacting case: mean-field solution}
\label{sec:pure-condensate}
We consider now the interacting solution of Eq.~(\ref{eq:quantum-langevin-Bose}) for the interacting case $g\neq 0$. As a first level of approximation, we consider the classical limit of  Eq.~(\ref{eq:quantum-langevin-Bose}), which, in absence of a reservoir, corresponds to the well-known Gross-Pitaevskii description of a condensate~\cite{Pitaevskii_book}.
This can be accomplished by replacing the quantum field $\hat{\psi}$ with a classical complex field $\psi$ and by neglecting the quantum noise $\hat{\xi}$. The classical field $\psi$ can be thus interpreted as the wave function of a condensate.

The validity of this approximation relies on the fact that the non-condensed fraction is assumed to be very small: this would have to be checked a posteriori by studying the effect of the fluctuations on the stability of the condensate solution (see Sec.~\ref{sec:quantum-fluctuations}). While in lower dimensional geometries, fluctuations are expected to be dominant \cite{Mora_quasi_BEC,Altman_2D,Sieberer_2D_vortices} and thus preclude any such description, we expect that for high enough spatial dimension $d$ condensation is possible \cite{Sieberer_3D,Sieberer_BEC}. Thus, a weak interaction coupling $g$ (inducing a weak quantum depletion), and a certain selectivity in frequency of the dissipation (limiting the generation of excitations of high energy)  should be suitable conditions for the emergence of coherence in the system. The classical field $\psi(\rr,t)$ thus obeys the following equation:
\begin{multline}
\label{eq:driven-GP}
\frac{\partial\psi(\rr,t)}{\partial t} = -i\left[\omega_0-\frac{\nabla^2}{2m}+g|\psi (\rr,t)|^2 \right]\psi(\rr,t)\\
+\int_\tau \Gamma(\tau)\psi(\rr,t-\tau),
\end{multline}
which has the form of a driven-dissipative Gross-Pitaevskii equation with a memory kernel. We focus on spatially homogeneous solutions of the form 
\begin{equation}
\label{eq:stationary-ansatz}
\psi(t) = \psi_0\,  e^{-i\omega_\text{BEC}t},
\end{equation}
which describe a condensate with infinite lifetime, frequency $\omega_\text{BEC}$ and density $n_0 = |\psi_0|^2$.

The non-condensed case $\psi_0 = 0$ is always a solution of Eq.~(\ref{eq:driven-GP}), whose stability may be studied by linearizing Eq.~(\ref{eq:driven-GP}) around it: this yields the linear equation studied in Sec.~\ref{sec:free-bose}. As a result, the non-condensed solution is stable when the spectrum of the excitations lies outside the amplifying region, i.e., $\omega_0+\epsilon_\kk \geq\omega_2$. We will now show that non-trivial, condensed ($\psi_0\neq 0$) solutions exist when the bare frequency lies below the upper-boundary of the amplifying region, i.e., $\omega_0 \leq \omega_2$. In this case, the interaction generates a blue-shift $\sim g n_0$ of the bosonic bare frequency $\omega_{0}$, thus providing a natural saturation mechanism as the condensate frequency is spontaneously set at one of the boundaries of the amplifying region. In fact, by inserting Eq.~(\ref{eq:stationary-ansatz}) into Eq.~(\ref{eq:driven-GP}), one finds
\begin{equation}
\label{eq:stationary-condition}
 \omega_\text{BEC} = \omega_0 +g |\psi_0|^2 +i \Gamma(\omega_\text{BEC})
\end{equation}
from which, by taking the real and the imaginary part and by using Eq.~(\ref{eq:KK-relations}), one finds the two following equations for $\omega_\text{BEC}$ and $|\psi_0|^2$:
\begin{subequations}
\label{eq:MF-equations}
\begin{align}
 \mathcal{S}_{\rm{p}}(\omega_{\text{BEC}}) &= S_{l}(\omega_{\text{BEC}}) \label{eq:pump-loss-balance}\\
 \omega_{\text{BEC}} &= \omega_0+\mu + \delta_\text{L}(\omega_\text{BEC}), \label{eq:condensate frequency}
\end{align}
\end{subequations}
where  
\begin{equation}
\mu\equiv g|\psi_0|^2
\end{equation}
is the mean-field self-interaction energy and
\begin{equation}
 \delta_\text{L}(\omega) = \text{PV}\int_{\omega'} \frac{1}{\omega-\omega'}\left[\mathcal{S}_{\rm{l}}(\omega')- \mathcal{S}_{\rm{p}}(\omega')\right]
\end{equation}
corresponds to a Lamb shift of the condensate frequency due to the contact with the bath.
From Eq.~(\ref{eq:pump-loss-balance}), we deduce that the only solutions for the condensate frequency are: $\omega_{\text{BEC}}=\omega_{1,2}$. However, the solution $\omega_1$ will be unstable, since the low energy excitations of the condensate will fall in the amplified region $[\omega_1,\omega_2]$ and undergo dynamical instability, thus we will not take into account this solution and consider in all the next sections the case $\omega_{\text{BEC}}=\omega_2$. 

We finally remark that, unlike usual VCSEL \cite{VCSEL} where stability is induced by a saturation effect of the pump (emitters are 'two-level like' nonlinear systems which need some time to be repumped in the excited state), stability is expected to be in our model a consequence of the interplay between  the frequency dependence of pumping and losses and the progressive blue-shift $g|\psi_0|^2$ induced by interactions during the condensate growth, this until the condensate frequency reaches $\omega_{\text{BEC}}$ where pump and losses perfectly compensate.

\subsection{Interacting case: Bogoliubov analysis of fluctuations}
\label{sec:quantum-fluctuations}

In order to study the stability of the condensate and to characterize the properties of its excitations, we express the bosonic field as
\begin{equation}
\label{eq:fluctuation-decomposition}
\hat{\psi}(\rr,t)=\left[\psi_0 + \hat{\Lambda}(\rr,t)\right]\ee^{-i\omega_{\text{BEC}}t},
\end{equation}
where $\hat{\Lambda}(\rr,t)$ is an operator describing the fluctuations above the condensate. Inserting this decomposition and the mean-field solution obtained from Eq.~(\ref{eq:MF-equations}) into Eq.~(\ref{eq:quantum-langevin-Bose}), and retaining terms up to the first order in the fields $\hat{\Lambda}(\rr)$, $\hat{\Lambda}^\dagger(\rr)$, one obtains 
\begin{equation}
\label{eq:linear-langevin}
\frac{\partial\hat{\Lambda}(\rr,t)}{\partial t}
=-i\com{\hat{\Lambda}(\rr,t)}{H_{\text{bog}}(t)}
+\int_\tau \widetilde{\Gamma}(\tau)\hat{\Lambda}(\rr,t-\tau)+\tilde{\xi}(\rr,t)
\end{equation}
where
\begin{multline}
\label{eq:bogoliubov-hamiltonian}
H_{\text{bog}}=\int \dd^d r \left\lbrace \hat{\Lambda}^{\dagger}(\rr)\frac{-\nabla^2}{2m} \hat{\Lambda}(\rr)+\frac{\mu}{2}\left[2\hat{\Lambda}^{\dagger}(\rr) \hat{\Lambda}(\rr)\right.\right.\\
\left.\left.+ \hat{\Lambda}(\rr) \hat{\Lambda}(\rr)+\hat{\Lambda}^{\dagger}(\rr)\hat{\Lambda}^{\dagger}(\rr)\right]\right\rbrace
\end{multline}
is the Bogoliubov Hamiltonian, $\widetilde{\Gamma}$ is defined as
\begin{equation}
\label{eq:Gamma-tilde}
\widetilde{\Gamma}(t)=\ee^{i\omega_{\text{BEC}}t}\Gamma(t) -\delta(t)\Gamma(\omega_{\text{BEC}}),
\end{equation}
and $\tilde{\xi}(\rr,t)=\ee^{i\omega_{\text{BEC}}t}\xi(\rr,t)$. After calculation of the commutator, the equation  Eq.~(\ref{eq:linear-langevin}) can be rewritten as
\begin{multline}
\label{eq:linear-langevin-comm}
\frac{\partial\hat{\Lambda}(\rr,t)}{\partial t}=-i\left\{\frac{-\nabla^2}{2m} \hat{\Lambda}(\rr,t)+\mu\left[ \hat{\Lambda}(\rr,t)+ \hat{\Lambda}^\dagger(\rr,t)\right]\right\}\\
+\int_\tau \widetilde{\Gamma}(\tau)\hat{\Lambda}(\rr,t-\tau)+\tilde{\xi}(\rr,t) .
\end{multline}
The linear system~(\ref{eq:linear-langevin-comm}) can be regarded as the driven-dissipative non-markovian counterpart of the Bogoliubov-de Gennes equations. Similarly to the equilibrium case, the field $\hat{\Lambda}(\rr,t)$ and its hermitian conjugate $\hat{\Lambda}^\dagger(\rr,t)$ are coupled by the interaction energy $\mu$: this coupling is mediated by processes in which non-condensed particles are scattered into the condensate, and vice-versa. It is convenient to rewrite Eq.~(\ref{eq:linear-langevin}) in momentum and frequency space: in order to do this, we define the Fourier transform of the fields and noise operators as in Eq.~(\ref{eq:Fourier-definitions}). The correlations of the quantum noise operators in the momentum and frequency space are given by:
\begin{subequations}
\label{eq:noise-correlation-frequency-momentum}
\begin{align}
\langle \tilde{\xi}_\kk(\omega)\tilde{\xi}^{\dagger}_{\kk '}(\omega')\rangle & = \delta_{\kk-\kk'}\, \delta_{\omega-\omega'}\mathcal{S}_{\rm{l}}(\omega_{\text{BEC}}+\omega), \\
\langle \tilde{\xi}^{\dagger}_{\kk}(\omega)\hat{\xi}_{\kk'}(\omega')\rangle & =\l\delta_{\kk-\kk'}\, \delta_{\omega-\omega'} \mathcal{S}_{\rm{p}}(\omega_{\text{BEC}}+\omega).
\end{align}
\end{subequations}
with $\delta_{\kk}\equiv \left(2\pi\right)^{d}\delta^{(d)} (\kk)$, $\delta_{\omega}\equiv 2\pi\delta(\omega)$.
After taking the Fourier transform of Eq.~(\ref{eq:linear-langevin}), we obtain the following set of coupled equations :
\begin{equation}
\label{eq:langevin-linear-system}
\omega\left(\begin{array}{c}
\hat{\Lambda}_\kk(\omega)\\
\hat{\Lambda}^{\dagger}_{-\kk}(-\omega)
\end{array}\right)
=\mathcal{L}_{\kk}(\omega)\left(\begin{array}{c}
\hat{\Lambda}_{\kk}(\omega)\\
\hat{\Lambda}^{\dagger}_{-\kk}(-\omega)
\end{array}\right)+i\left(\begin{array}{c}
\tilde{\xi}_{\kk}(\omega)\\
\tilde{\xi}^\dagger_{-\kk}(-\omega)
\end{array}\right),
\end{equation}
where the matrix $\mathcal{L}_{\kk}(\omega)$ is given by
\begin{equation}
\label{eq:langevin-matrix}
\mathcal{L}_{\kk}(\omega)=
\left(\begin{array}{ccc}
\epsilon_k+\mu+i\widetilde{\Gamma}(\omega) & \mu \\
-\mu& -\epsilon_k-\mu+i\widetilde{\Gamma}^*(-\omega) \\
\end{array} \right),
\end{equation}
where $\widetilde{\Gamma}(\omega)$ is the Fourier transform of $\widetilde{\Gamma}(t)$ defined in Eq.~(\ref{eq:Gamma-tilde}), and it reads:
\begin{equation}
\label{eq:gamma-tilde-frequency}
\widetilde{\Gamma}(\omega)=\Gamma(\omega+\omega_{\text{BEC}})-\Gamma(\omega_{\text{BEC}}),
\end{equation}
and we used the notation $\widetilde{\Gamma}^*(\omega)\equiv [\widetilde{\Gamma}(\omega)]^* $.  The complex function $\widetilde{\Gamma}(\omega)$ represents the frequency-dependent decay rate (real part) and lamb shift (imaginary part) of the fluctuations. $\widetilde{\Gamma}(\omega)$ vanishes for $\omega\to 0$, consistently with the fact that the condensate has an infinite lifetime (see Eq.~(\ref{eq:stationary-condition})).

For later convenience, we define the correlation matrix $\mathcal{C}_{\kk}(\omega)$
\begin{multline}
\label{eq:correlation-matrix}
\delta_{\kk-\kk'}\, \delta_{\omega-\omega'}\, \mathcal{C}_{\kk}(\omega)=\\
\left(\begin{array}{cc}
\langle\hat{\Lambda}_\kk(\omega)\hat{\Lambda}^\dagger_{\kk'}(\omega^{'})\rangle & \langle \hat{\Lambda}_\kk(\omega)\hat{\Lambda}_{-\kk'}(-\omega{'})\rangle \\
\langle\hat{\Lambda}^{\dagger}_{-\kk}(-\omega)\hat{\Lambda}^{\dagger}_{\kk'}(\omega^{'})\rangle & \langle\hat{\Lambda}^{\dagger}_{-\kk}(-\omega)\hat{\Lambda}_{-\kk'}(-\omega^{'})\rangle
\end{array}\right),
\end{multline}
which can be calculated by inverting Eq.~(\ref{eq:langevin-linear-system}), multiplying the solution by its hermitian conjugate and averaging over the noise correlation using Eq.~(\ref{eq:noise-correlation-frequency-momentum}) (see App.~\ref{app:quantum-correlations} for the details of the calculations).
\begin{figure}[t]
\includegraphics[width=0.99\columnwidth]{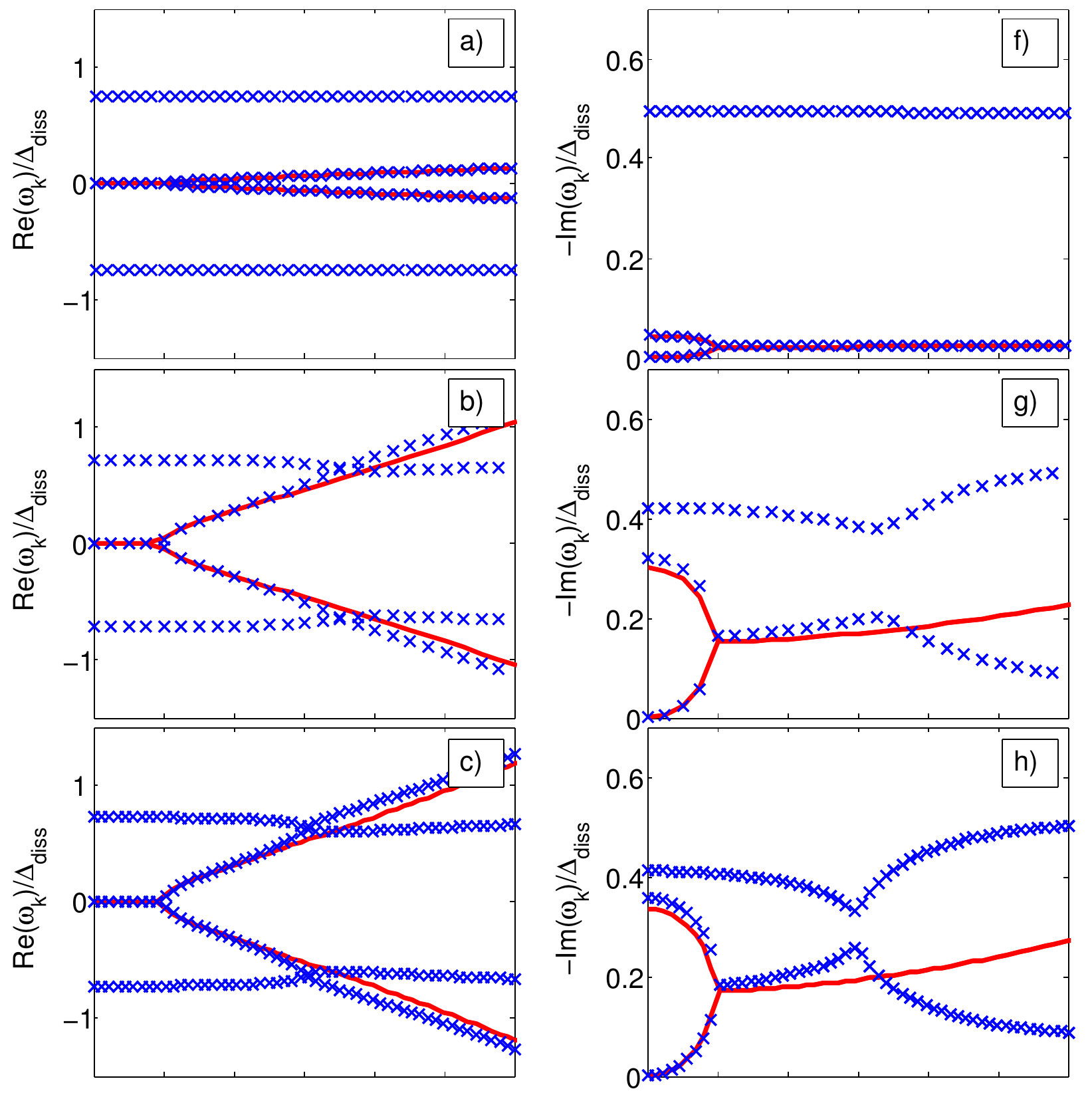}\vspace{-2.2mm}\\
\includegraphics[width=0.998\columnwidth]{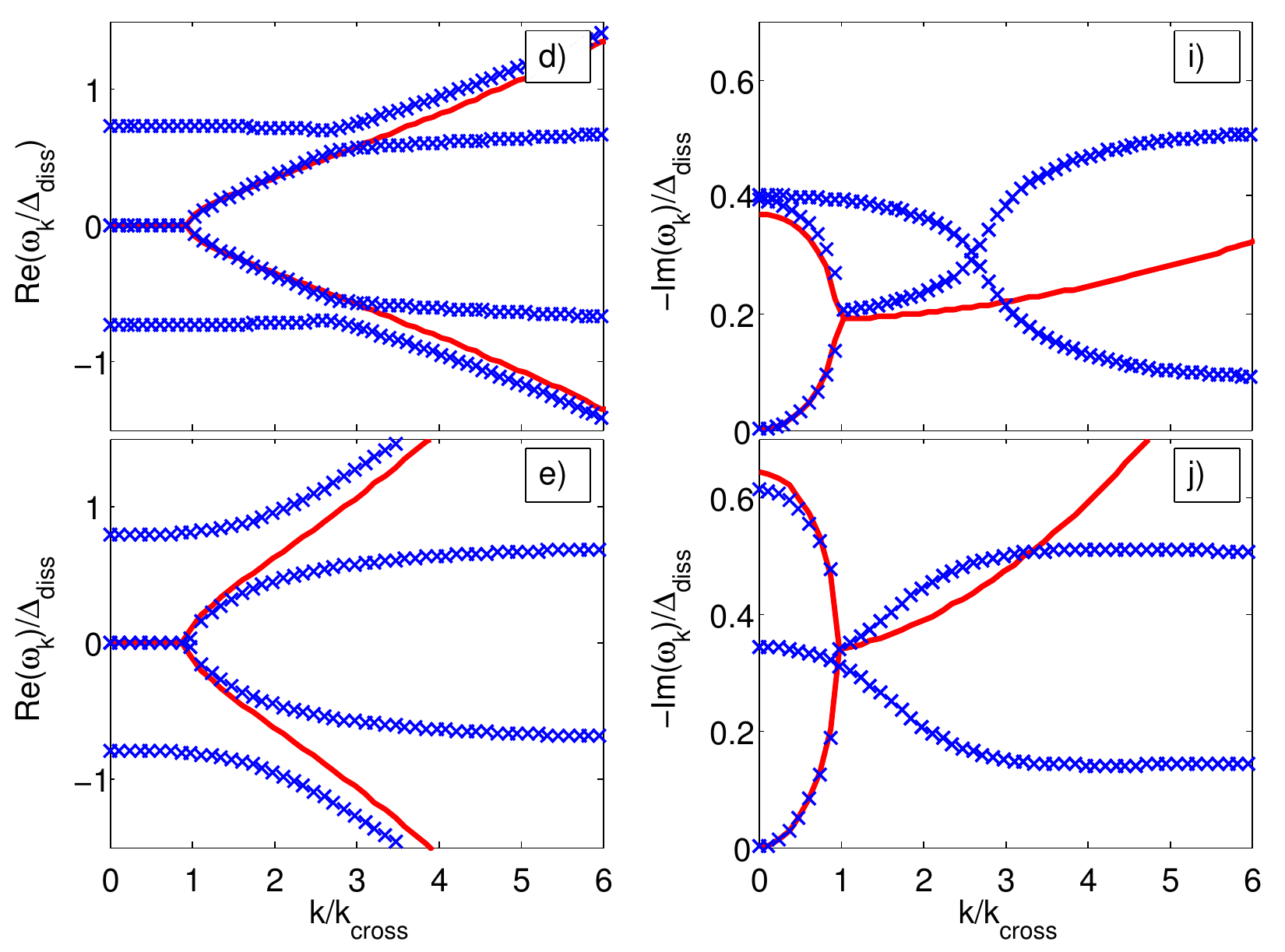}
\caption{\label{fig:modes}
Excitation spectrum of the condensate in the case a Lorentzian pump spectrum and Markovian losses (model defined in Sec.\ref{sec:model}). Left (resp. right) panel: real (resp. imaginary) part of the frequency in units of $\Delta_{\text{diss}}$  in function of the momentum $k$ in units of $k_{cross}$ defined as $|z_R|E_{k_{cross}}=z_I\mu$. In blue crosses we plot exact numerical values for the eigenfrequencies $\omega_{\kk,n}$ of the full non-Markovian theory (Eq.~\ref{eq:langevin-linear-system})), and in red solid lines the solutions $\omega_\kk^\pm$ given by the corresponding Markovian effective theory at low energies (Eq.~(\ref{eq:mode-frequencies})). Going from upper to lower panels, we investigate the transition between weak-dissipation to strong-dissipation. Parameters: $m=1$, $\delta/\Delta_{\text{diss}}=0$, $\Gamma_{l}/\Gamma_{\rm{p}}=0.3$. From up to down, $\Gamma_{\rm{p}}^0/\Delta_{\rm{diss}}=0.1,\, 0.55,\, 0.6,\, 0.65,\, 1$. 
}
\end{figure}
\subsection{Dynamical stability of excitations}\label{eq:dynamical-stability}
In order to study the dynamical stability of the mean-field solution, it is necessary to check that the elementary excitations do not grow exponentially and have a finite lifetime. To this end, we derive from Eq.~(\ref{eq:langevin-linear-system}) the excitations spectrum by calculating frequencies $\omega_{\kk,n}$ (with $i$ some integer number used to label the excitation) which cancel out the determinant of the matrix $\omega-\mathcal{L}_{\kk}(\omega)$ with $\mathcal{L}_{\kk}(\omega)$ defined in Eq.~(\ref{eq:langevin-matrix}). This leads us to the following condition on the frequency:
\begin{equation}
\label{eq:stability-equation}
\left[ \omega - \epsilon_k - \mu - i\widetilde{\Gamma}(\omega) \right] \left[ \omega + \epsilon_k + \mu - i\widetilde{\Gamma}^*(-\omega) \right] + \mu^2 = 0.
\end{equation}
Solutions with negative imaginary parts correspond to decaying excitations, while in presence of any instability, some solutions present a positive imaginary part. Since we are considering generic non-Markovian systems, $\tilde{\Gamma}(\omega)$ can be any function verifying the Kramers-Kronig relations reported in Eq.~(\ref{eq:KK-relations}), thus in general Eq.~(\ref{eq:stability-equation}) may have a large number of solutions, and it may be not possible to solve it analytically.

In the case of Markovian losses and a Lorentzian spectrum Eq.~(\ref{eq:lorentzian-Markovian-power-spectra}), Eq.~(\ref{eq:stability-equation}) becomes an algebraic equation which admits four different solutions, thus giving rise to four different branches by varying the momentum $k$ which we computed numerically. In Fig.~\ref{fig:modes}, these solutions are plotted successively for increased values of $\Gamma_{(\text{l}/\text{p})}$, going at fixed ratio $\Gamma_{\rm{l}}/\Gamma_{\rm{p}} = 0.3$ from a weakly-dissipative regime  (upper panels) in which the spectral power $\Gamma_{(\text{l}/\text{p})}$ are weak with respect to the linewidth $\Delta_{\text{diss}}$, to a strongly-dissipative regime (lower panels) in which they become comparable or higher. All other parameters (interaction $g$, mass $m$, detuning $\delta$, linewidth $\Delta_{\text{diss}}$) are left unchanged.

As a first observation, all imaginary parts of the frequencies are negative, so there is no instability (we checked this for other choice of parameters). Secondly, in the weakly-dissipative regime (panels a) and f)) the mode structure is typical of exciton-polariton driven-dissipative condensates \cite{Szymanska_BEC_polariton,Wouters_excitations,Carusotto_rev,Chiocchetta_Langevin} and presents a sharp transition from purely damped modes to propagating ones.  Also we observe two other branches of imaginary part $\Delta_{\text{diss}}$ and real parts $\pm (\omega_{BEC}-\omega_{\rm{p}})$: these additional frequencies account for the oscillation of bath degrees of freedom, which are hidden in the non-Markovianity of the Langevin equation and are nearly unaffected by the system dynamics due to the weak coupling (In a photonic language for the Lorentzian pump spectrum, the reservoir degrees of freedom responsible for the photonic pumping may be seen as two-level emitters of transition frequency $\omega_{\rm{p}}$). 

However, for stronger dissipation (other panels), the system and reservoir degrees of freedom are coupled and can not be treated separately, which can be seen in a clearest way by a deformation of the various branches near the crossing point. Remarkably, a sharp transition from weak to strong coupling occurs between the panels c),h) and the panels d),i), inducing a change in excitation spectrum structure, as one moves from a situation of branch crossing to an avoided crossing: in this regime, the collective modes associated with the excitation spectrum couples the bosonic and the bath degrees of freedom, giving birth to a mixed quasi-excitation. In a photonic language, this suggests that some elementary excitations are of a polaritonic nature.
\subsection{Effective low-frequency Markovian dynamics}
\label{sec:low-frequency-dynamics}
Here we show that it is possible to derive an effective time-local equation describing the dynamics for frequencies small enough with respect to  $\Delta_{\rm{diss}}$: indeed, for $\omega \ll \Delta_\text{diss}$, the function $\widetilde{\Gamma}(\omega)$ defined in Eq.~(\ref{eq:langevin-linear-system}) can be linearized and approximated as $\widetilde{\Gamma}(\omega) \approx \omega \widetilde{\Gamma}'(0)=\omega \Gamma' ( \omega_\text{ BEC } )$.  As a result, the low-frequency limit of the Langevin equation Eq.~(\ref{eq:langevin-linear-system}) becomes: 
\begin{equation}
\label{eq:low-energy-langevin-frequency}
\omega
\hat{\Lambda}_\kk(\omega)
= z\left\{\epsilon_k\hat{\Lambda}_\kk(\omega)+\mu\left[\hat{\Lambda}_\kk(\omega)+\hat{\Lambda}_{-\kk}^\dagger(-\omega)\right]+i\overline{\xi}_{\kk}(\omega)\right\},
\end{equation}
with the coefficient $z$ defined as
\begin{equation}
\label{eq:z}
z = \lim_{\omega\to 0}\left[\frac{\omega}{\omega - i\widetilde{\Gamma}(\omega)}\right] = \left[ 1-
i \Gamma' ( \omega_\text{ BEC } ) \right]^{-1},
\end{equation}
and the new noise operators $\overline{\xi}_{\kk}(\omega)$ and $\overline{\xi}_{\kk}^\dagger(\omega)$ are characterized by the correlations
\begin{subequations}
\begin{align}
\langle \overline{\xi}_\kk(\omega)\overline{\xi}^{\dagger}_{\kk '}(\omega')\rangle &=\delta_{\kk-\kk'}\, \delta_{\omega-\omega'}\mathcal{S}_{\rm{l}}(\omega_{\text{BEC}}), \\
\langle \overline{\xi}^{\dagger}_{\kk}(\omega)\overline{\xi}_{\kk'}(\omega')\rangle &=\delta_{\kk-\kk'}\, \delta_{\omega-\omega'}\mathcal{S}_{\rm{p}}(\omega_{\text{BEC}}).
\end{align}
\end{subequations}
Notice that the noise operators $\overline{\xi}_{\kk}(\omega)$ and $\overline{\xi}_{\kk}^\dagger(\omega)$ correspond to an effective classical noise, since their correlations do not depend on the order of the operators, as a consequence of Eq.~(\ref{eq:pump-loss-balance}).

With respect to a purely hamiltonian dynamics, all couplings in the commutator have been multiplied by the complex number $z$. The eigenmodes of Eq.~(\ref{eq:low-energy-langevin-frequency}) are given by  
\begin{equation}
\label{eq:mode-frequencies}
\omega^{\pm}_k = - i z_I \left( \epsilon_k + \mu \right) \pm \sqrt{ z_R^2 E_k^2 - z_I^2 \mu^2 }, 
\end{equation}
where $z=z_R-i z_I$, $z_R$ and $z_I$ are both real numbers, and $E_k = \sqrt{\epsilon_k(\epsilon_k+2\mu)}$ is the equilibrium Bogoliubov energy for the Hamiltonian Eq.~(\ref{eq:bogoliubov-hamiltonian}). We can already verify the dynamical instability of the mean-field solution for the choice of BEC frequency $\omega_{\rm{BEC}}=\omega_{1}$, as this leads to a negative $ z_I$ (due to a change of sign in the derivative of the real part of $\Gamma(\omega)$  involved in Eq.~(\ref{eq:z})) and thus to a positive imaginary part in the low-momentum excitation spectrum in Eq.~(\ref{eq:mode-frequencies}). This justifies definitively the choice $\omega_{\rm{BEC}}=\omega_2$ (whose dynamical stability was already checked in \ref{eq:dynamical-stability}).

The frequencies $\omega_k^\pm$, shown in Fig.~\ref{fig:modes} in red solid lines, closely resemble the spectrum of a polaritonic driven-dissipative condensate~\cite{Szymanska_BEC_polariton,Wouters_excitations,Carusotto_rev,Chiocchetta_Langevin}: they are are imaginary for small momenta, which signals the purely diffusive nature of low-energy excitations, while they acquire a finite real part at higher momenta. In particular, for $k\to 0$ the branch $\omega_k^+$ vanishes and therefore it can be identified with the (diffusive) Goldstone mode associated with the spontaneous breaking of the $U(1)$ symmetry. As was already discussed in the previous subsection, higher powers of $\omega$ present in Eq.~(\ref{eq:langevin-linear-system}) related to the non-Markovianity can generate additional modes not predicted by the effective low-energy theory Eq.~(\ref{eq:low-energy-langevin-frequency}), which can be observed in Fig.~\ref{fig:modes}. 

The validity of Eq.~(\ref{eq:low-energy-langevin-frequency}) for the study of the long-range physics has to be checked a posteriori, by requiring the absolute value $|\omega_{k}^{\pm}|$ to be small with respect to $\Delta_\text{diss}$ for small $k$, so that it can be computed by mean of the low-energy effective theory Eq.~(\ref{eq:low-energy-langevin-frequency}). On the one hand, this condition is naturally satisfied for the Goldstone branch $\omega_k^+$ for low enough momenta. On the other hand, the gapped branch $\omega_k^-$ verifies $|\omega_{k=0}^{-}|=2 z_I \mu$, and therefore the gapped mode is correctly described by the Markovian low-frequency theory only if $2 z_I \mu\ll \Delta_\text{diss}$. According to Eq.~(\ref{eq:z}), $z$ scales as $S_{l}(\omega_{BEC})/\Delta_{\text{diss}}$, so the gapped mode is correctly described by the Markovian low-frequency theory only if $S_{l}(\omega_{BEC})\mu\ll \Delta_{\text{diss}}^2$: this is the case for very small power spectra (weak dissipation) or very small interaction energy $\mu$. The validity of this analysis is illustrated in the panels a) and f) of Fig.\ref{fig:modes}, which feature the case of a weak dissipation, and where we can see that the theoretical prediction Eq.~(\ref{eq:mode-frequencies}) for the Goldstone mode and the gapped mode accurately fits with the exact numerical predictions.

\begin{figure}[t]
\centering
\includegraphics[width=1\columnwidth,clip]{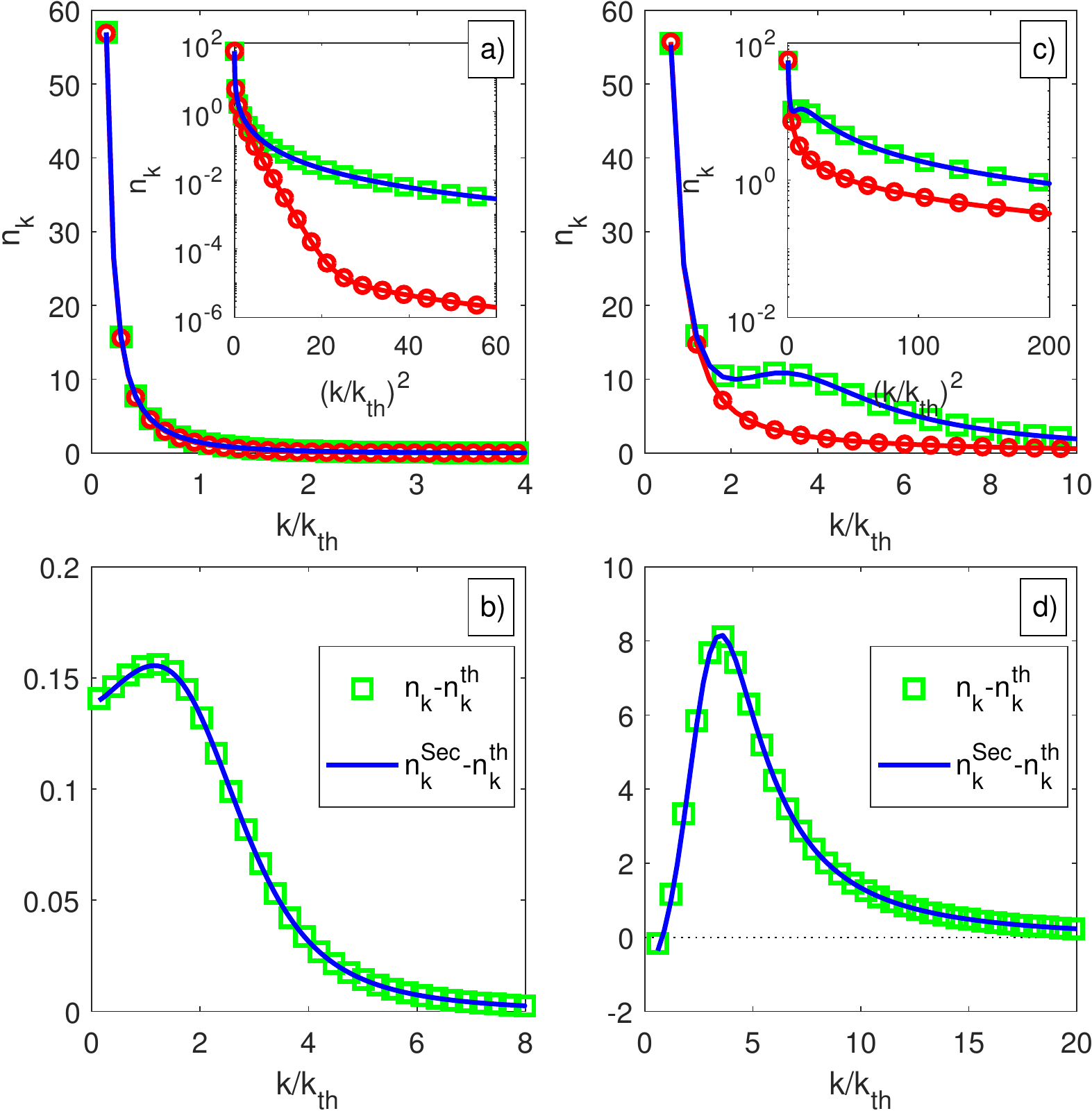}
\caption{\label{fig:static-weak}Static properties of the condensate at steady state in the weakly-dissipative regime ( i.e., for the loss and pump power spectra $\Gamma_{\rm{l}}$ and $\Gamma_{\rm{p}}$ much smaller than the reservoirs' characteristic spectral width $\Delta_{\text{diss}}$) in the case of Lorentzian  pump power spectrum and Markovian losses (model defined in Sec.\ref{sec:model}). The left (resp. right) panels correspond to a detuning between the cavity and the atoms chosen to induce a weak (resp. strong) chemical potential $\mu$ with respect to the effective temperature $ T_{\text{eff}}$. Upper panels: static correlations $n_\kk=\langle \Lambda_{\kk}^{\dagger}\Lambda_{\kk}\rangle$ in function of the momentum $\kk$ in units of $\kk_{\text{th}}$ defined by $E(\kk_{\text{th}})= T_{\text{eff}}$, and in inset, their logarithm in function of the square momentum $\kk^2$ in units of $\kk_{\text{th}}^2$. In green squares we plot the steady state properties given by numerical calculations of the linearized Langevin equation (Eq.~(\ref{eq:langevin-linear-system})) in the weakly-dissipative regime, in red lines with circles the results given by the Grand-Canonical ensemble (Eq.~(\ref{eq:thermal-static-correlations})), and in dashed blue lines the analytical results given by the secular approximation (Eq.~(\ref{eq:secular-static-correlations})). Lower panels: the absolute error $n_k-n_k^{\rm{th}}$ in green squares lines (resp. $n_k^{{\text{Sec}}}-n_k^{\rm{th}}$ in dashed blue lines) between the numerical solution of the Langevin equation (resp. the analytical solution given by the secular approximation) and the thermal case, in function of the momentum $\kk$ in units of $\kk_{\text{th}}$. Parameters: for all panels, $m=1$, $\Gamma_{l}/\Gamma_{\rm{p}}^0=0.3$,  $\Gamma_{\rm{p}}/\Delta_{\text{diss}}=10^{-2}$. Deduced quantity  $T_{\text{eff}}/\Delta_{\text{diss}}=0.55$. For the left (resp. right) panels: $\delta/\Delta_{\text{diss}}=0.72$ (resp. $-10$). Deduced quantity $\mu/\Delta_{\text{diss}}=4.6\times 10^{-2}$ (resp. $10.8\times 10^0$).}
\end{figure}
\section{Pseudo-thermalization}
\label{sec:effective-thermalization}
In this section we give evidence for low-energy pseudo-thermalization for generic power spectra, both at static and dynamical level, by showing that the low-energy static correlations map on equilibrium ones, and demonstrating the validity of the FDT in the the low frequency regime. We also compute analytically the static correlations at all energies in the weakly-dissipative regime. Finally, in the specific choice of reservoirs where the Kennard-Stepanov relation is exactly verified, we demonstrate the validity of the FDT at all frequencies, and show that the steady-state in the weakly-dissipative regime is in a Gibbs ensemble.
\subsection{Static correlations}
\label{sec:static-correlations}
The steady-state properties of a system undergoing low-energy pseudo-thermalization should look like those of a Gibbs ensemble at low-energies. In Sec.\ref{sec:low-energy-satic-correlation}, we give the low-energy analytical expression for static correlations, both in the weakly and strongly-dissipative regimes, while in Sec.~\ref{sec:ansatz-steady-state} we give an exact analytical expression at all energies, only valid in the weakly-dissipative regime.

\subsubsection{Low energies}
\label{sec:low-energy-satic-correlation}

In this section we focus on the low-energy regime $E_{\rm{k}}\ll\Delta_{\text{diss}}$. By using the expressions derived in App.~\ref{app:quantum-correlations} for the frequency-correlation matrix $\mathcal{C}_{k}(\omega)$ defined in Eq.~(\ref{eq:correlation-matrix}) and by restricting ourselves to the low-frequency regime using the procedure described in  Sec.~\ref{sec:low-frequency-dynamics},  we compute by Fourier transform the steady state values of the momentum distribution $n_{k}=\langle\hat{\Lambda}^\dagger_\kk\hat{\Lambda}_{\kk}\rangle $ and the anomaleous average $\mathcal{A}_{k}=\langle\hat{\Lambda}_\kk\hat{\Lambda}_{-\kk}\rangle $ at leading order in $E_{\rm{k}}/\Delta_{\rm{diss}}$ (see App.~\ref{app:static-correlations} for the details of the calculation):
\begin{eqnarray}
\label{eq:effective-thermal-static-correlations}
n_{k}&\simeq&\frac{T_\text{eff}\,(\epsilon_k +\mu)}{\left(E_k\right)^{2}},\\
\label{eq:effective-thermal-static-correlations-anomaleous}
\mathcal{A}_{k}&\simeq &-\frac{T_\text{eff}\:\mu}{\left(E_k\right)^{2}},
\end{eqnarray}
where we remind that  $T_{\text{eff}}$ is defined in Eq.~(\ref{eq:effective-temperature}). These static correlations have to be compared to those obtained by doing a Bogoliubov calculation for a Bose gas at thermal equilibrium of temperature $T_\text{eff}$ and chemical potential $\mu=g|\psi_0|^2$:
\begin{eqnarray}
\label{eq:thermal-static-correlations}
n^{\text{th}}_{k}&=&\frac{1}{e^{\beta_\text{eff}E_\kk}-1}(|u_k|^2+|v_k|^2)+|v_k|^2\\
&\underset{(\beta_\text{eff} E_{\kk})\to 0}{\simeq}&\frac{T_\text{eff}\,(\epsilon_k +\mu)}{\left(E_k\right)^{2}},\nonumber\\
\label{eq:thermal-static-correlations-anomaleous}
\mathcal{A}^{\text{th}}_{k}&=&2\left(\frac{1}{e^{\beta_\text{eff}E_k}-1}+\frac{1}{2}\right) u_k v_k^*, \\
&\underset{(\beta_\text{eff} E_{k})\to 0}{\simeq}&-\frac{T_\text{eff}\:\mu}{\left(E_k\right)^{2}},\nonumber
\end{eqnarray}
where $u_{k}$ and $v_{k}$ relate the annihilation operator $\hat{\Lambda}_{\kk}$ to the phonon annihilation (resp. creation) operator $\hat{b}_\kk$ (resp. $\hat{b}_\kk^\dagger$)  through the Bogoliubov transformation:
\begin{eqnarray}
\label{eq:equilibrium-Bogoliubov-transformation}
\hat{\Lambda}_k&=&u_k \hat{b}_\kk+v_k^* \hat{b}_\kk^\dagger,\\
u_k &=&\frac{1}{2}\left[\sqrt{\frac{\epsilon_k}{E_k}}+\sqrt{\frac{E_k}{\epsilon_k}}\right],\\
v_k &=&\frac{1}{2}\left[\sqrt{\frac{\epsilon_k}{E_k}}-\sqrt{\frac{E_k}{\epsilon_k}}\right].
\end{eqnarray}
By comparing Eqs~(\ref{eq:effective-thermal-static-correlations}),(\ref{eq:effective-thermal-static-correlations-anomaleous}) and Eqs.~(\ref{eq:thermal-static-correlations}),(\ref{eq:thermal-static-correlations-anomaleous}), we note that the low-energy limit $\beta_\text{eff} E_{k}\to 0$ of  the driven-dissipative quantum Langevin model accurately reproduce a thermal infrared behaviour, leading to the so-called Rayleygh-Jeans distribution. {Strikingly the validity of this equilibrium signature only depends on the condition $E_{k}\ll\Delta_{\text{diss}}\propto T_{\text{eff}}$, and in particular is not restricted to the range of Bogoliubov energies $E_{k}$ below the interaction energy $\mu$: in the regime $T_{\text{eff}}\gg\mu$, one expects thus the full phonon-particle crossover in the elementary excitations to be well represented by an equilibrium  theory.} Correlations at higher energies $E_{k}\geq\Delta_{\text{diss}}\propto T_{\text{eff}}$ are not expected although to be thermal: in particular we do not expect necessarily to see exponential tails.

The analytical arguments leading to the expressions Eqs.~(\ref{eq:effective-thermal-static-correlations}),(\ref{eq:effective-thermal-static-correlations-anomaleous}) can be verified in Fig.\ref{fig:static-weak} (resp. Fig.~~\ref{fig:static-strong}), where we plot the static correlations obtained by numerical resolution of the linearized Langevin equation (\ref{eq:linear-langevin-comm}) for a Markovian loss spectrum and Lorentzian pump spectrum (Eq.~(\ref{eq:lorentzian-Markovian-power-spectra})),  in the weakly-dissipative regime (resp. strongly-dissipative regime), i.e, for $\Gamma_{\rm{p}}^0,\,\Gamma_{l}\ll \Delta_{\text{diss}}$ (resp.  $\Gamma_{\text{p}}^0,\,\Gamma_{l}$ of the order of $\Delta_{\text{diss}}$), and compare those correlations to thermal ones. We plotted the static correlations for two detunings $\delta$ of the bare frequency $\omega_{0}$ with respect to the pump resonance $\omega_{\rm{p}}$, inducing different effective chemical potentials $\mu$, which is a decreasing function of $\delta$. Indeed, looking at Eq.~(\ref{eq:condensate frequency}) and neglecting as a first step the Lamb shift, we see that increasing the frequency of the pump $\omega_{\rm{p}}$ defined in Eq.~(\ref{eq:lorentzian-Markovian-power-spectra}), i.e., diminuishing the detuning $\delta=\omega_{0}-\omega_{\rm{p}}$ at fixed $\omega_{0}$, has for effect to increase $\omega_{\text{BEC}}$, and thus to increase also the chemical potential $\mu$.  

The case of a chemical potential weak (resp. strong) with respect to the effective temperature $T_{\text{eff}}$ is plotted in the left (resp. right) panels. The upper panels correspond to the static correlations (with in insets their logarithm to check for any high-energy exponential tails),  while in the lower panels we plot the absolute error $n_k-n_k^{\rm{th}}$ between the solutions of the Langevin equations with respect to thermal predictions. Expectedly, static correlations given by the numerical simulation of the Langevin equation (green squares) coincide with the equilibrium results (red solid line with circles) at energies lower than the temperature (since $T_{\rm{eff}}$ scales as the spectra linewidth $\Delta_{\rm{diss}}$ and is of the same order of magnitude), both in the weakly- and strongly-dissipative regimes. In particular, they diverge as $1/k^2$ at low momenta, and looking at the absolute errors we note the that the corresponding corrections to thermal equilibrium remain finite at low energies and thus surprisingly do not present any subsingular divergencies $\propto 1/k$, so effective thermal equilibrium seems also to be true also at the next leading order at a static level for this particular system. 

However, as we expected, the pseudo-thermalization does not extend for a generic choice of power spectra at higher energy scales (see the logarithmic plot) as the Kennard-Stepanov relation is not valid in this energy range: in particular, while one can see in the Grand Canonical distribution the presence of exponential tails of a Boltzmann type in the panel a) of Fig.\ref{fig:static-weak} (approximately for momenta verifying $2\leq k^2/k_{\rm{th}}^2\leq 25$, the slower decay for higher momenta being related to the dominant vacuum fluctuations), such behaviour is not present in the driven-dissipative steady-state which rather features algebraic decay. This feature is specifically related to the Lorentzian shape for the pump spectrum Eq.~(\ref{eq:lorentzian-Markovian-power-spectra}) chosen for numerical simulations. In the case of a big chemical potential $\mu>T_{\text{eff}}$ (see Fig.\ref{fig:static-weak}~[panel c)]), the thermal distribution does not present exponential tails neither because the vacuum fluctuations which decay algebraically are dominant with respect to thermal fluctuation in the energy range $E_k\ge T_{\rm{eff}}$. 

\begin{figure}[t]
\centering
\includegraphics[width=1\columnwidth,clip]{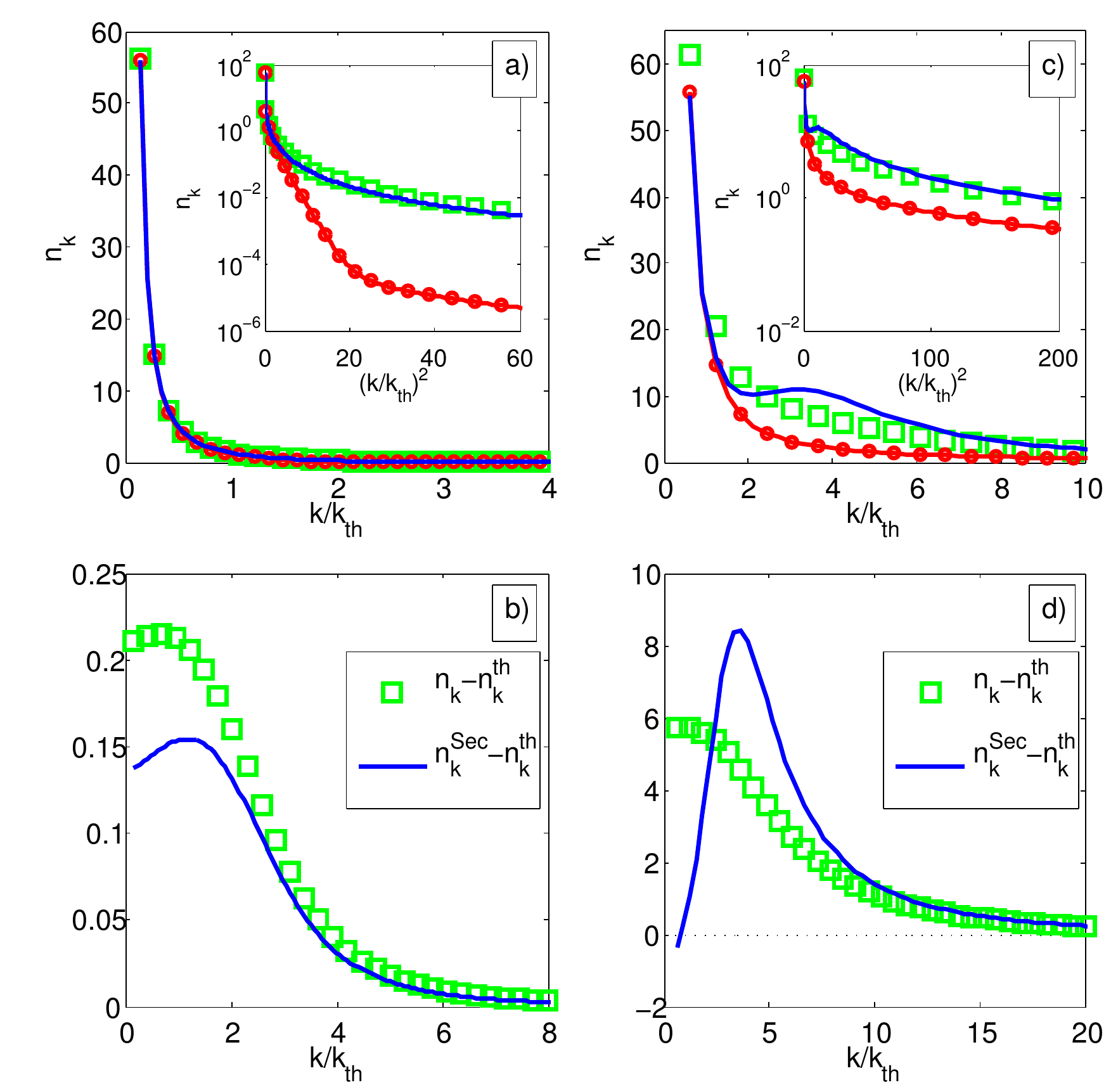}
\caption{\label{fig:static-strong}Static properties of the condensate at steady state in the strongly-dissipative regime (i.e., for the loss and pump power spectra $\Gamma_{\rm{l}}$ and $\Gamma_{\rm{p}}$ comparable to the reservoirs characteristic spectral width $\Delta_{\text{diss}}$)  in the case of Lorentzian pump spectrum and Markovian losses (model defined in Sec.\ref{sec:model}). The left (resp. right) panels correspond to a detuning between the cavity and the atoms chosen to induce a weak (resp. strong) chemical potential $\mu$ with respect to the effective temperature $ T_{\text{eff}}$. Upper panels: static correlations $n_k=\langle \Lambda_{\kk}^{\dagger}\Lambda_{\kk}\rangle$ in function of the momentum $\kk$ in units of $\kk_{\text{th}}$ defined by $E(k_{\text{th}})= T_{\text{eff}}$, and in inset, their logarithm in function of the square momentum $\kk^2$ in units of $\kk_{\text{th}}^2$. In green squares we plot the steady state properties given by numerical calculations of the linearized Langevin equation (Eq.~(\ref{eq:langevin-linear-system})) in the strongly-dissipative regime, in red line with circles the results given by the Grand-Canonical ensemble (Eq.~(\ref{eq:thermal-static-correlations})), and in dashed blue lines the analytical results given by the secular approximation (Eq.~(\ref{eq:secular-static-correlations})). Lower panels: the absolute error $n_k-n_k^{\rm{th}}$ in green squares lines (resp. $n_k^{{\text{Sec}}}-n_k^{\rm{th}}$ in dashed blue lines) between the numerical solution of the Langevin equation (resp. the analytical solution given by the secular approximation) and the thermal case, in function of the momentum $\kk$ in units of $\kk_{\text{th}}$. Parameters: for all panels, $m=1$, $\Gamma_{l}/\Gamma_{\rm{p}}=0.3$,  $\Gamma_{\rm{p}}^0/\Delta_{\text{diss}}=1$. Deduced quantity  $T_{\text{eff}}/\Delta_{\text{diss}}=0.55$. For the left (resp. right) panels: $\delta/\Delta_{\text{diss}}=0.92$ (resp. $-10$). Deduced quantity $\mu/\Delta_{\text{diss}}=7.3\times 10^{-2}$ (resp. $11.0\times 10^0$).}
\end{figure}
\subsubsection{Analytical expressions for the static correlations at all energies in the weakly-dissipative regime\label{sec:ansatz-steady-state}}
When the dissipation strength $\mathcal{S}_{(\rm{l}/\rm{p})}(\omega)$ is much weaker than the linewidth of the power spectra $\Delta_{\text{diss}}$, it is possible to provide exact analytical predictions for the static correlations at all momenta:
\begin{eqnarray}
\label{eq:secular-static-correlations}
n^{\text{Sec}}_{k}&=&\frac{1}{K(E_{k})-1}(|u_k|^2+|v_k|^2)+|v_k|^2,\\
\label{eq:secular-static-correlations-anomaleous}
\mathcal{A}^{{\text{Sec}}}_{k}&=&\left(\frac{1}{K(E_{k})-1}+\frac{1}{2}\right) u_k v_k^*.
\end{eqnarray}
Comparing these expressions to Eqs.~(\ref{eq:thermal-static-correlations}),(\ref{eq:thermal-static-correlations-anomaleous}), we see that the vacuum properties are left unchanged with respect to equilibrium statistics, while the Boltzmann factor $\ee^{\beta E_{k}}$ of the Bose-Einstein distribution for phononic excitations in the Grand canonical ensemble has been replaced by the non-equilibrium factor: 
\begin{equation}
\label{eq:non-eq-boltzmann}
K(E_{k})=\frac{\mathcal{S}_{l}(\omega_{\text{BEC}}+E_{k})|u_{k}|^2+\mathcal{S}_{\rm{p}}(\omega_{\text{BEC}}-E_{k})|v_{k}|^2}{\mathcal{S}_{\rm{p}}(\omega_{\text{BEC}}+E_k)|u_{k}|^2+\mathcal{S}_{l}(\omega_{\text{BEC}}-E_{k})|v_{k}|^2},
\end{equation}
giving thus rise to the modified Bose Einstein phonon distribution $1/[K(E_{k})-1]$. 

The factor $K(E_{k})$ can be interpreted as the ratio between the annihilation and creation rates (both induced by pumping and losses dissipative processes) of a single phononic excitation at the Bogoliubov energy $E_{k}$, and is calculated using the secular approximation (valid in the weakly-dissipative regime). The phonon distribution and average occupation number are a consequence of an emerging detailed balance between states with $N_\kk$ and $N_\kk-1$ phonons of momentum $\kk$. 

We note that if the pumping and loss rates verify the Kennard-Stepanov condition Eq.~(\ref{eq:exact-KS}), one recovers the equilibrium Boltzmann factor $K(E_{k})=\ee^{\beta E_{k}}$: as expected the system is fully thermal at all energies, and its density matrix at steady-state is a Grand-Canonical ensemble. In the general case by using Eqs.~(\ref{eq:effective-temperature})-(\ref{eq:approximate-KS}) we note that $K(E_{k})= 1+\beta_{\text{eff}} E_{k}+\mathcal{O}\left(E_{k}/\Delta_{\rm{diss}}\right)^2\sim \ee^{\beta E_{k}}$ for $E_{k}/\Delta_{\rm{diss}}\to0$:  this provides us another confirmation that low-energy static properties should be thermal.

The static correlations computed under the secular approximation expressed in Eqs.~(\ref{eq:secular-static-correlations}),(\ref{eq:secular-static-correlations-anomaleous})are shown in dashed blue lines in the upper panels of Fig.~\ref{fig:static-weak} (resp. Fig.~\ref{fig:static-strong}) and compared with the exact numerical results obtained from the linearized Langevin equation (\ref{eq:linear-langevin-comm}) in the weakly- (resp. strongly-) dissipative regime. In the lower panels we plot the absolute error $n_k^{\text{Sec}}-n_k^{\rm{th}}$ between the solution given by the secular approximation and the thermal distribution. In the weakly-dissipative regime we note absolutely no difference between the exact numerical solution and $n_k^{\text{Sec}}$. Expectedly, in the strongly-dissipative regime they coincide only at low momenta ($E_{k}\ll T_{\text{eff}}$) (up to a finite error, which is small with respect to the divergency in $1/k^2$), and do not provide exact results at higher momenta. The accuracy at low-energies of Eqs.~(\ref{eq:secular-static-correlations}),(\ref{eq:secular-static-correlations-anomaleous}) also in the strongly-dissipative regime stems from the fact that low-energy pseudo-thermalization is true in both the weakly- and strongly-dissipative regimes, as shown in the previous subsection.

We now justify the expression Eqs.~(\ref{eq:secular-static-correlations}),(\ref{eq:secular-static-correlations-anomaleous}) for the static correlations in the weakly-dissipative regime $\mathcal{S}_{\rm{p}},\mathcal{S}_{l}\ll\Delta_{\text{diss}}$: in such a secular regime, dissipation can be considered as a "classical" stochastic process inducing transitions in the system $S$ between the eigenstates of the Bogoliubov hamiltonian $H_{\text{bog}}$ defined in Eq.~(\ref{eq:bogoliubov-hamiltonian}). These eigenstates  are labelled by the phononic occupancy number: $\otimes_{\kk}\ket{N_{\kk}}$. Here $\kk$ is the momentum and $N_{\kk}$ is the occupation number of the phonon of momentum $\kk$. The phonon annihilation and creation operators $\hat{b}_{\kk}$ and $\hat{b}_{\kk}^\dagger$ are related to the particle annihilation  and creation operators $\hat{\Lambda}_{\kk}$ and $\hat{\Lambda}_{\kk}^\dagger$ by the Bogoliubov transformation Eq.~(\ref{eq:equilibrium-Bogoliubov-transformation}).

\textit{Phonon annihilation rate:}
Let us calculate as a first step the phononic annihilation rate. Starting from a state with $N_{\kk}$ phonons of momentum $\kk$ and Bogoliubov energy energy $E_k$, one can remove one phonon through two processes:
\begin{itemize}
\item First, one can remove a phonon by losing a particle of momentum $\kk$. The total energy removed to the system is $\omega_{\text{BEC}}+E_k$. This leads to the partial rate:
\begin{eqnarray}
\mathcal{T}^{(l)}(N_{\kk}\rightarrow N_{\kk}-1)&=&\mathcal{S}_{l}(\omega_{\text{BEC}}+E_k)\left|\bra{N_\kk-1}\hat{\Lambda}_{\kk}\ket{N_\kk}\right|^2\nonumber\\
&=&\mathcal{S}_{l}(\omega_{\text{BEC}}+E_k)N_{\kk}|u_k|^2.
\end{eqnarray}
Starting from a wave-function calculation, this expression could have been alternatively recovered by mean of the Fermi's Golden rule \cite{Grynberg_book}.
\item  However, due to the presence of counter-rotating terms in the Bogoliubov theory, it is also possible to remove a phonon by pumping a particle of momentum $-\kk$. The total energy added to the system in that case is $\omega_{\text{BEC}}-E_k$, i.e, the mean-field energy of a single photon, minus the energy of the phonon excitation. Thus the corresponding rate is:
\begin{eqnarray}
\mathcal{T}^{(p)}(N_{\kk}\rightarrow N_{\kk}-1)&=&\mathcal{S}_{\rm{p}}(\omega_{\text{BEC}}-E_k)\nonumber\\
&&\phantom{\mathcal{S}\omega_{\rm{BEC}}}\times\left|\bra{N_\kk-1}\hat{\Lambda}_{\kk}^\dagger \ket{N_\kk}\right|^2\nonumber\\
&=&\mathcal{S}_{\rm{p}}(\omega_{\text{BEC}}-E_k)N_{\kk}|v_k|^2
\end{eqnarray}
\end{itemize}
The total phonon loss rate is thus:
\begin{multline}
\mathcal{T}^{(tot)}(N_{\kk}\rightarrow N_{\kk}-1)=\mathcal{S}_{l}(\omega_{\text{BEC}}+E_k)N_{\kk}|u_k|^2\\
+\mathcal{S}_{\rm{p}}(\omega_{\text{BEC}}-E_k)N_{\kk}|v_k|^2.
\end{multline}

\textit{Phonon creation rate:}
One can calculate similarly the phonon total creation rate. Starting from a state with $N_k-1$ phonons of momentum $\kk$ and Bogoliubov energy energy $E_k$, one can add one phonon by pumping a new particle (the total energy added to the system is thus $\omega_{\text{BEC}}+E_k$) or by losing a particle (the total energy lost is $\omega_{\text{BEC}}-E_k$). After a calculation very similar to the previous paragraph, one obtains the following expression:
\begin{multline}
\mathcal{T}^{(tot)}(N_{\kk}-1\rightarrow N_{\kk})=\mathcal{S}_{\rm{p}}(\omega_{\text{BEC}}+E_k)N_{\kk}|u_k|^2\\
+\mathcal{S}_{l}(\omega_{\text{BEC}}-E_k)N_{\kk}|v_k|^2.
\end{multline}

\textit{Phonon probability distribution:}
The ratio between the phonon annihilation and creation rates is given by
\begin{eqnarray}
K(E_{k})&=&\frac{\mathcal{T}^{(tot)}(N_{\kk}\rightarrow N_{\kk}-1)}{\mathcal{T}^{(tot)}(N_{\kk}-1\rightarrow N_{\kk})}\\
&&=\frac{\mathcal{S}_{l}(\omega_{\text{BEC}}+E_{k})|u_{k}|^2+\mathcal{S}_{\rm{p}}(\omega_{\text{BEC}}-E_{k})|v_{k}|^2}{\mathcal{S}_{\rm{p}}(\omega_{\text{BEC}}+E_k)|u_{k}|^2+\mathcal{S}_{l}(\omega_{\text{BEC}}-E_{k})|v_{k}|^2}.\nonumber
\end{eqnarray}
Because dissipative processes can remove or add only one phonon of momentum $\kk$ at a time and can not affect simultaneously the phononic occupancy at other momenta, one deduces that at steady state the probabilities $\pi(...,N_\kk-1,...)$ and $\pi(...,N_\kk,...)$ of having $N_\kk-1$ and $N_\kk$ phonons of momentum $\kk$ verify the following detailed balance relation :
\begin{equation}
\pi(N_\kk-1)=K(E_{k})\pi(N_\kk).
\end{equation}
One deduces that the probability distribution is 
\begin{equation}
\pi(N_\kk)=\frac{1}{1-K(E_k)^{-1}} K(E_{k})^{-n},
\end{equation}
and that the average phonon occupation number is
\begin{equation}
n^{\text{Sec},\text{phon}}_{\kk}=\frac{1}{K(E_k)-1}.
\end{equation}
Doing a Bogoliubov transformation Eq.~(\ref{eq:equilibrium-Bogoliubov-transformation}), one obtains the static momentum  distribution and anomaleous averages Eqs.~(\ref{eq:secular-static-correlations}),(\ref{eq:secular-static-correlations-anomaleous}).
\subsection{Effective temperature from the FDT}
\label{sec:FDT}
\begin{figure}[t]
\centering
\includegraphics[width=1\columnwidth,clip]{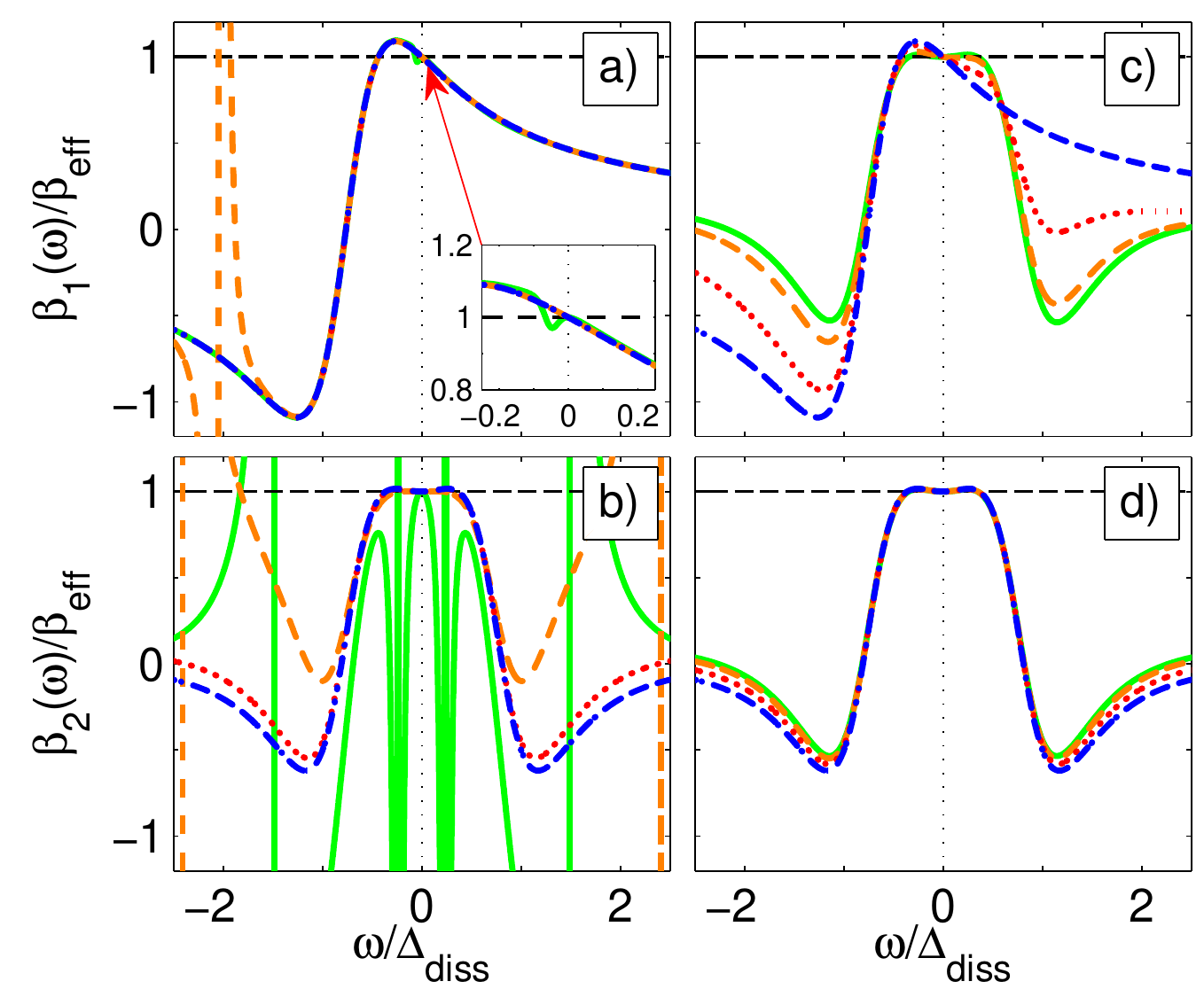}
\caption{\label{fig:FDT}Test of the FDT/KMS relations for various sets of parameters. Upper (resp. lower) panels: plot of the frequency dependent effective temperature $\beta_{1}(\kk,\omega)$ (resp. $\beta_{2}(\kk,\omega)$)  defined in Eq.~(\ref{eq:effective-temperature}) for a Lorentzian pump and Markovian losses, in function of the frequency $\omega$ in units of $\Delta_{\text{diss}}$, and for various momenta $\kk$.  Panels a),b) (resp. c),d)) use the same parameters as in the panels a),b) (resp. c),d)) of Fig.~\ref{fig:static-weak}. For each panel, the various curves correspond to increasing values of the momentum $k$, chosen in such a way that the corresponding Bogoliubov energies span a wide energy range across the effective temperature $T_{\rm{eff}}=0.54\Delta_{\rm{diss}}$: $k/k_{\rm{th}}=0.18$ for the green solid line, $k/k_{\rm{th}}=3.65$, for the orange dashed line, $k/k_{\rm{th}}=9.1$ for the red dotted line, $k/k_{\rm{th}}=54.7$ for the blue dash-dotted line}
\end{figure}
\begin{figure}[t]
\centering
\includegraphics[width=0.8\columnwidth,clip]{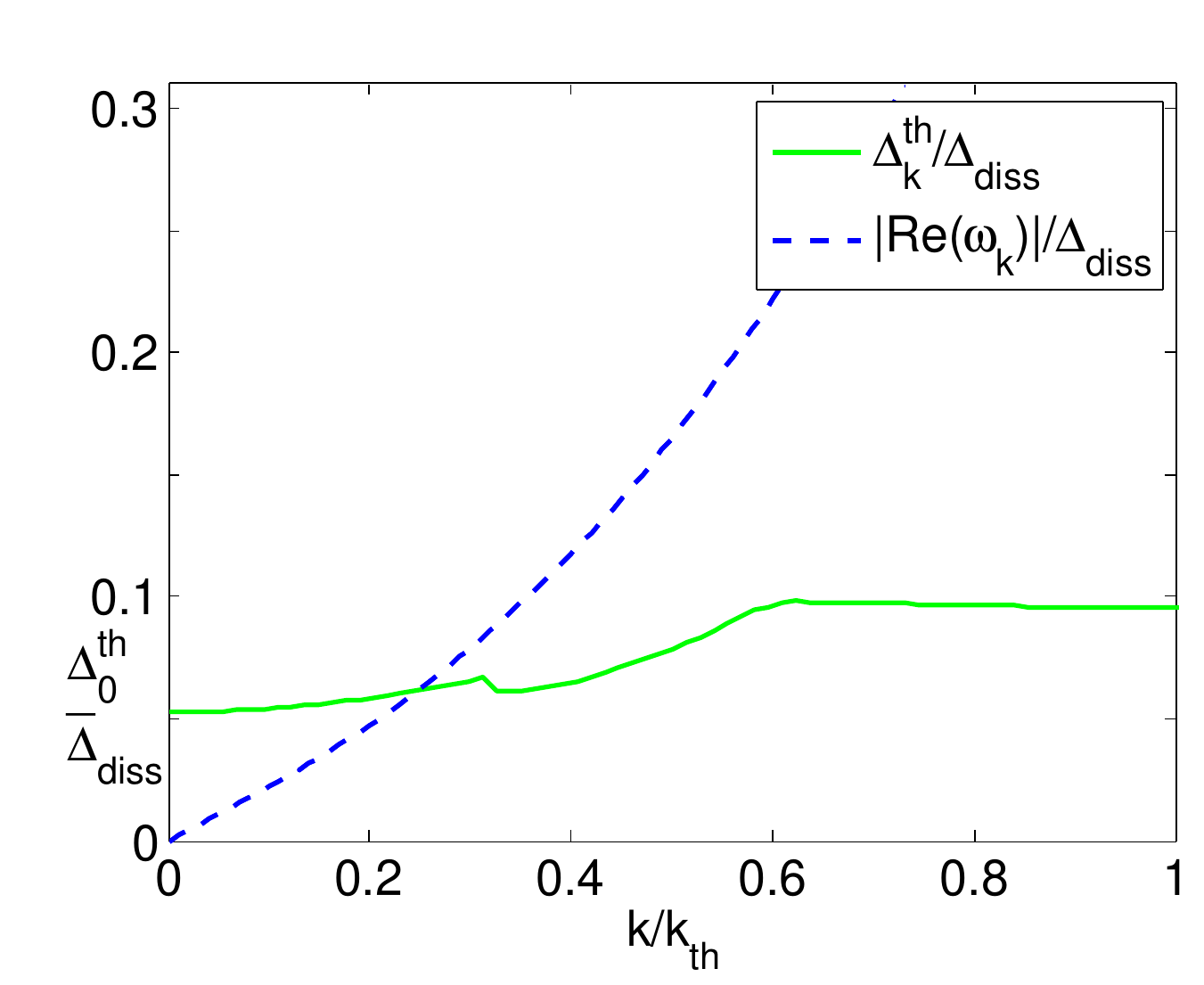}
\caption{\label{fig:efficiency}Test of the efficiency of thermalization in function of the momentum $k$ in units of $k_{\rm{th}}$. In green solid line, one shows $\Delta^{\rm{th}}_{\kk}$ (in units of $\Delta_{\rm{diss}}$), defined as the maximum frequency such that both conditions $|\beta_{1}(\kk,\omega)-\beta_{\rm{eff}}|/\beta_{\rm{eff}}\leq 0.05$ and  $|\beta_{2}(\kk,\omega)-\beta_{\rm{eff}}|/\beta_{\rm{eff}}\leq 0.05$ are verified for all $\omega$ contained in the interval $|\omega-\omega_{\rm{BEC}}|\leq \Delta^{\rm{th}}_{\kk}$. In dashed blue lines, one shows in absolute value the real part $|\text{Re}(\omega^i_k)|$ (in units of $\Delta_{\rm{diss}}$) of the dissipative Bogoliubov spectrum for the same parameters, computed by exact numerical calculation of the solutions of Eq.~(\ref{eq:stability-equation}) (the other branches are not visible here since located at higher energies). Same parameters as in Fig.~\ref{fig:FDT} a),b)}
\end{figure}
A remarkable consequence of equilibrium involving dynamical quantities is the so-called fluctuation-dissipation theorem~\cite{Kubo_FDT}, which provides a relationship between the linear response of a system to an external perturbation and the correlation of thermal fluctuations. 

Let us define the symmetrized correlation ($C$) and response ($R$) functions for two arbitrary operators $\hat{A}$ and $\hat{B}$ as
\begin{subequations}
\label{eq:greens-functions}
\begin{align}
iC(t-t') & = \langle \{ \hat{A}(t), \hat{B}(t')\}\rangle, \\
iR(t-t') & = \theta(t-t')\langle [ \hat{A}(t), \hat{B}(t')]\rangle,
\end{align}
\end{subequations}
where the time dependence of $\hat{A}(t)$ and $\hat{B}(t)$ is determined in the Heisenberg picture, while the average $\langle\dots \rangle$ is taken over an equilibrium state at temperature $T$. As a consequence of equilibrium, $C$ and $R$ depend only on the time difference $t-t'$ and therefore we can define their Fourier transforms $C(\omega)/R(\omega)= \int_t\ee^{i\omega t}C(t)/R(t)$. The explicit form of the FDT then reads:
\begin{equation}
\label{eq:FDT}
C(\omega) = 2\coth\left(\beta \omega/2\right) \text{Im}[R(\omega)],
\end{equation}
with $\beta = T^{-1}$.
An alternative, fully equivalent formulation of the FDT is the so-called Kubo-Martin-Schwinger (KMS) \cite{Kubo_KMS,Martin_Schwinger_KMS} condition: 
\begin{align}
\label{eq:KMS}
S_{AB}(-\omega) = \ee^{-\beta \omega} S_{BA}(\omega),
\end{align}
where ${S}_{AB}(t) = \langle \hat{A}(t) \hat{B} \rangle$ and ${S}_{BA}(t) = \langle \hat{B}(t) \hat{A} \rangle$.

The FDT and KMS condition have often been used as a tool to probe the actual thermalization in classical and quantum systems, and to characterize the eventual departure from equilibrium \cite{Cugliandolo_effective_temperature,Foini_effective_temperature,Chiocchetta_thermalization}. In particular, from Eqs.~(\ref{eq:FDT}) and~(\ref{eq:KMS}) one can define an effective frequency-dependent temperature $T_{A,B,\text{eff}}(\omega)$ such that the FDT or KMS condition are satisfied: if the system is really at equilibrium, then $T_{A,B,\text{eff}}(\omega)$ has a constant value $T$ which corresponds to the thermodynamic temperature. On the other hand, if the system is out of equilibrium it will generically develop a non-trivial dependence on $A$, $B$ and $\omega$.

In the following, we discuss the effective temperatures obtained from the linearized equation Eq.~(\ref{eq:linear-langevin}): in this respect, we will consider the following ratios: 
\begin{align}
\frac{\langle \hat{\Lambda}_\kk(\omega)\hat{\Lambda}^\dagger_{\kk}\rangle}{\langle\hat{\Lambda}^\dagger_\kk(\omega)\hat{\Lambda}_\kk\rangle} & =
\frac{\mathcal{S}_{\rm{l}}(\omega_\text{BEC}+\omega)+\mathcal{S}_{\rm{p}}(\omega_\text{BEC}-\omega)A_k(\omega)}{\mathcal{S}_{\rm{p}}(\omega_\text{BEC}+\omega)+\mathcal{S}_{\rm{l}}(\omega_\text{BEC}-\omega)A_k(\omega)},\label{eq:FDT-ratio-1} \\
\frac{\langle \hat{\Lambda}_\kk(\omega)\hat{\Lambda}_{-\kk}\rangle}{\langle\hat{\Lambda}_{\kk}(-\omega)\hat{\Lambda}_{-\kk}\rangle} & =
\frac{\mathcal{S}_{\rm{l}}(\omega_\text{BEC}+\omega)+\mathcal{S}_{\rm{p}}(\omega_\text{BEC}-\omega)B_k(\omega)}{\mathcal{S}_{\rm{p}}(\omega_\text{BEC}+\omega)+\mathcal{S}_{\rm{l}}(\omega_\text{BEC}-\omega)B_k(\omega)},\label{eq:FDT-ratio-2}
\end{align}
where the functions $A_k(\omega)$ and $B_k(\omega)$ are explicitly reported in App.~\ref{app:quantum-correlations}.
At thermal equilibrium, the value of the ratios~(\ref{eq:FDT-ratio-1}) and~(\ref{eq:FDT-ratio-2}) is fixed by Eq.~(\ref{eq:KMS}) while, in the present case, they have a nontrivial dependence on $\omega$ and $k$, since the system is out of equilibrium.  

We then define the effective (inverse) temperatures
\begin{align}
\label{eq:effective-temperatures}
\beta_{1}(\kk,\omega) & = \frac{\dd}{\dd\omega}\log\left[\frac{\langle \hat{\Lambda}_\kk(\omega)\hat{\Lambda}^\dagger_{\kk}\rangle}{\langle\hat{\Lambda}^\dagger_\kk(\omega)\hat{\Lambda}_\kk\rangle} \right], \\
\label{eq:effective-temperatures_anomaleous}
\beta_{2}(\kk,\omega) & =\frac{\dd}{\dd\omega} \log\left[  \frac{\langle \hat{\Lambda}_\kk(\omega)\hat{\Lambda}_{-\kk}\rangle}{\langle\hat{\Lambda}_{\kk}(-\omega)\hat{\Lambda}_{-\kk}\rangle}\right],
\end{align}
which are generic functions of $k$ and $\omega$, and can be evaluated by using Eqs.~(\ref{eq:FDT-ratio-1}) and (\ref{eq:FDT-ratio-2}). However, inserting the functional forms Eqs.~(\ref{eq:FDT-ratio-1}),~(\ref{eq:FDT-ratio-2}) into Eqs.~(\ref{eq:effective-temperatures}),~(\ref{eq:effective-temperatures_anomaleous}) we see that for $\omega\to 0$, both $\beta_{1}(\kk,\omega)$ and $\beta_{2}(\kk,\omega)$ tend toward the same $\kk$-independent value $\beta_\text{eff}$ defined in Eq.~(\ref{eq:effective-temperature}), indicating that the KMS condition and the FDT are asymptotically verified at low frequencies. 

Remarkably, if the system satisfies the Kennard-Stepanov relation
\begin{equation}
\label{eq:detailed-balance}
 \mathcal{S}_{\rm{p}}(\omega_\text{BEC}+\omega)=\mathcal{S}_{\rm{l}}(\omega_\text{BEC}+\omega)e^{-\beta\omega},
\end{equation} 
then $\beta_{1}(\kk,\omega)=\beta_{2}(\kk,\omega) = \beta$ for every value of $\omega$ and $\kk$, i.e., the system is at full thermal equilibrium, even if the environment is highly non-thermal (see Secs.~\ref{subsec:artificial-Kennard-Stepanov},~\ref{subsec:excitons} for examples of physical systems made of non-thermal reservoirs verifying artificially the KS relation).

In the left (resp. right) panels of Fig.~\ref{fig:FDT}, we plot the effective temperature $\beta_{1}(\kk,\omega)$ (resp. $\beta_{2}(\kk,\omega)$) as a function of $\omega$ in units of $\Delta_{\text{diss}}$, for various values of the momentum $k$. On one hand, in the region $\omega\ll \Delta_{\text{diss}}$, these effective temperatures converge to the same value $\beta_{\text{eff}}$. This demonstrates the low-frequency validity of the FDT and confirms that the system is effectively thermalized in that frequency range. Even though these plots focus on the weakly-dissipative regime, the same behaviour was also found in the strongly-dissipative regime, which displays identical features. On the other hand, away from the low-frequency region the effective temperatures have a non-trivial, frequency-momentum dependent behaviour, so the system is globally not at equilibrium.

In order to conclude from these plots that the Bogoliubov modes are actually thermalized, one should check that the low-energy limit value of $\beta_{\text{eff}}$ is already (approximately) attained by $\beta_{1,2}(\kk,\omega)$ at the frequency $\omega_{\kk,n}$ of the mode. In other terms, one needs to verify that the strong modulations that one sees in Fig.~\ref{fig:FDT} are located at energies above the mode frequency. To put this reasoning on quantitative grounds, we can define an energy cutoff $\Delta^{\rm{th}}_k$ as the maximum frequency such that the conditions $|\beta_{1,2}(\kk,\omega)-\beta_{\rm{eff}}|/\beta_{\rm{eff}}\leq \epsilon$ are verified for all $\omega$ contained in the interval $|\omega-\omega_{\rm{BEC}}|\leq \Delta^{\rm{th}}_{\kk}$. This sets a quantitative criterion for thermalization, which of course depends on the value of the small parameter $\epsilon$. In practice, we shall adopt $\epsilon=0.05$. The most constraining condition is the one on $\beta_2(\omega)$ in the $k\to 0$ limit, that sets $\Delta^{\rm{th}}_0\simeq 0.051\times\Delta_{\rm{diss}}\simeq 0.1\times T_{\rm{eff}}$: 
we have checked that for no value of $k$ the peaks of $\beta_{2}(\omega)$ can get any closer to $\omega=0$. Note that the peaks are not actual singularities for a finite dissipation, still they get sharper and sharper in the limit of a weak dissipation.

By comparing the green and black lines in Fig.~\ref{fig:efficiency}, one sees that all the low-energy elementary excitations of the condensate have their resonance located in the thermalized frequency window $[-\Delta^{\rm{th}}_0,\Delta^{\rm{th}}_0]$ and will verify the FDT at a very good level of approximation. This is a strong evidence of their effective thermalization. Remarkably, for this simulation the energy cutoff $\Delta^{\rm{th}}_0$ is slightly bigger than the effective chemical potential $\mu\simeq 0.045\times\Delta_{\rm{diss}}$, meaning that not only the phononic region of the spectrum is efficiently thermalized, but also part of the crossover to the single-particle regime.


\section{Derivation of the Langevin equation from a quantum optics microscopic model\label{sec:Langevin-derivation}} 
In this section, we proceed to the derivation of the Langevin equation (\ref{eq:quantum-langevin-Bose}) in an lattice geometry, starting from the microscopic quantum optics model introduced in \cite{Lebreuilly_2016}. 
Namely, we consider a photonic driven-dissipative Bose-Hubbard lattice made of $L$ nonlinear cavities coupled by tunneling. Each cavity possesses a natural frequency $\omega_{0}$ and is assumed to contain a $\chi^{(3)}$ Kerr nonlinear medium, which induces effective repulsive interactions between photons lying in the same cavity. Dissipative phenomena due finite mirror transparency and absorption by the cavity material are responsible for (possibly non-Markovian) loss processes. 

We assume that a large number $N_{\rm{at}}$ of two-level atoms are embedded in each cavity and that their transition frequencies $\omega_{\rm{at}}^{(n)}$ are distributed according to the distribution $\mathcal{D}(\omega)$. Each atom is coupled to the cavity with a Rabi frequency $\Omega_{\text{R}}$ and is incoherently pumped into its excited state at a fast rate $\Gamma_{\rm{p}}^{\rm{at}}$ so that spontaneous decay can be neglected. The small value of the individual Rabi coupling $\Omega_{\text{R}}$ is compensated by the large number of atoms, which allows for a non-negligible and controllable collective coupling to the photonic cavity modes, whereas having $\Gamma_{\rm{p}}^{\rm{at}} \gg \Omega_{\text{R}}$ guarantees that each atom spends most of its time in its excited state.


The whole system dynamics can be described by an Hamiltonian involving the photonic and atomic degrees of freedom plus an external environment (modelled as a series of baths of harmonic oscillators):
\begin{equation}
H=H_{\rm{ph}}+H_{\rm{at}}+H_{\text{I}}+H_{\text{bath}}+H_{\text{I},\text{bath}}.
\end{equation}
The Hamiltonian for the isolated photonic system has the usual Bose-Hubbard form 
\begin{equation}
H_{\rm{ph}}=\sum_{i=1}^{L}\left[\omega_{0}a_{i}^{\dagger}a_{i}+\frac{U}{2}a_{i}^{\dagger}a_{i}^{\dagger}a_{i}a_{i}\right]-\sum_{\avg{i,j}}\left[\hbar Ja_{i}^{\dagger}a_{j}+hc\right],\label{eq:phys-system-hamiltonian}
\end{equation}
where we assumed that the Kerr nonlinearity of the cavity medium induces an on-site interaction term $U$.
The free evolution of the atoms and their coupling to the photonic degrees of freedoms are described by the following terms
\begin{equation}
H_{\rm{at}}=\sum_{i=1}^{L}\sum_{n=1}^{N_{\rm{at}}}\omega_{\rm{at}}^{(n)}\sigma_{i}^{(n)+}\sigma_{i}^{(n)-}. \label{eq:phys-system-atoms}
\end{equation} 
and
\begin{equation}
H_{\text{I}}=\Omega_{\text{R}}\sum_{i,n}\left[a_{i}^{\dagger}\sigma_{i}^{-(n)}+hc\right],
\end{equation}
where the indices $i$ and $n$ account respectively for the lattice sites and the atoms in each site. 

Likewise the external environment and its coupling to the photonic and atomic degrees of freedom are represented by the following Hamiltonian contributions
\begin{equation}
H_{\text{bath}}=\sum_{i=1}^{L}\sum_{m}\left[\omega_{m}b_{i}^{(m)\dagger}b_{i}^{(m)}-
\sum_{n=1}^{N_{\rm{at}}}\tilde{\omega}_{m}c_{i}^{(n,m)\dagger}c_{i}^{(n,m)}\right],\label{eq:phys-system-bath}
\end{equation}
and
\begin{multline}
H_{\text{I},\text{bath}}=\sum_{i,m}g_m\left[a_{i}^{\dagger}b_{i}^{(m)}+hc\right]\\
+\sum_{i,n,m}\tilde{g}_m\left[\sigma_{i}^{+(n)} c_{i}^{\dagger(n,m)}+hc\right],\label{eq:phys-system-interaction}
\end{multline}
where the indices $m$ account for the various bath excitations. 

Remarkably, while the photonic field $a_{i}^{(m)}$ is coupled to the bath by mean of a creation operator $b_{i}^{(m)\dagger}$ with a positive frequency $\omega_{n}$ in order to account for loss processes such as radiative losses, the atomic raising operator $\sigma_{i}^{+(n)}$ is coupled in an anti-rotating way to a creation operator $c_{i}^{\dagger(n,m)}$ with a negative frequency $-\tilde{\omega}_{m}$ so to reproduce the effect of an irreversible atomic pumping leading to an inversion of population. In different terms, this process can be seen as the result of a negative temperature, as the atomic environment is more likely to induce an increase in energy than to have a cooling impact. Physically, such dissipative amplification effect can be reproduced in analogy with the lasing operation \cite{Scully_book} by coherently coupling the atomic ground-state to an additional third atomic level with a strong decay toward the first excited state. 

We assume both baths to be in the vacuum state at the initial time
\begin{equation}
\left\langle b_{i}^{(m)\dagger}b_{i}^{(m)}\right\rangle(0)=\left\langle c_{i}^{(n,m)\dagger}c_{i}^{(n,m)}\right\rangle (0)=0:
\end{equation}
meaning that the bath $b_{i}^{(m)}$ (resp. $c_{i}^{(n,m)}$) can only induce photon losses (resp. atomic excitation). The various baths are also assumed to have a broad spectral function
\begin{eqnarray}
\sum_m |g_m|^2 e^{-i\omega_{m}\tau}&=&\int_{\omega}\mathcal{S}_{\rm{l}}(\omega)e^{-i\omega\tau}\label{eq:phys-system-bath-loss-spectral-density}\\
\sum_m |\tilde{g}_m|^2 e^{-i\tilde{\omega}_{m}\tau}&=&\Gamma_{\rm{p}}^{\rm{at}}\delta(\tau),\label{eq:phys-system-bath-pump-spectral-density}
\end{eqnarray}
where $\mathcal{S}_{\rm{l}}(\omega)$ is the loss power spectra of a single cavity, and the atomic pumping processes are described as Markovian.

A consequence of being in the regime $\Gamma_{\rm{p}}\gg\Omega_{\text{R}}$ is that a single atom will have a very weak probability to be in the ground-state and that the effect of atomic saturation on photonic emission process will be strongly suppressed. We can thus model atoms as linear degrees of freedom, and replace the spin matrix of each atomic two-level system by an `inverse' harmonic oscillator whose vacuum state (resp. whose state with a single excitation) corresponds to the atomic excited state (resp. to the atomic ground-state): $\sigma_{i}^{(n)+}\Rightarrow a_{\text{at},i}^{(n)}$. States of the harmonic oscillator with more than one excitation will be so rarely occupied that they will not contribute to the photonic dynamics.

We obtain thus the modified (although physically equivalent) Hamiltonians contribution involving atomic degrees of freedom:
\begin{equation}
H_{\rm{at}}=\sum_{i=1}^{N_{cav}}\sum_{k=1}^{N_{\rm{at}}}(-\omega_{\rm{at}}^{(n)})a_{\text{at},i}^{(n),\dagger}a_{\text{at},i}^{(n)}+E_0
\label{eq:linearization-atom}
\end{equation} 
where $E_0$ is a constant, 
\begin{equation}
H_{\text{I}}=\Omega_{\text{R}}\sum_{i,n}\left[a_{i}^{\dagger}a_{\text{at},i}^{(n),\dagger}+hc\right], \label{eq:linearization-interaction}
\end{equation}
and
\begin{multline}
H_{\text{I},\text{bath}}=\sum_{i,m}g_m\left[a_{i}b_{i}^{(m)\dagger}+hc\right]+\\
\sum_{i,n,m}\tilde{g}_m\left[c_{i}^{(n,m)\dagger}a_{\text{at},i}^{(n)}+hc\right]. \label{eq:linearization-interaction-bath}
\end{multline}
Within this linearized form for the atomic dynamics, it is possible to derive an exact non-Markovian Langevin equation for the photonic quantum field, by reexpressing the Hamiltonian dynamics into the form of Heisenberg equations of motion for the various operators :
\begin{eqnarray}
\partial_{t}a_{i}(t)&=&-i\com{a_{i}(t)}{H_{\rm{ph}}(t)},\label{eq:heisenberg-photon}\\
&&-i\sum_{m}g_m^* \hat{b}_{i}^{(m)}(t)-i\Omega_{\text{R}}\sum_{k} a_{\text{at},i}^{(n)\dagger}(t)\nonumber\\
\partial_{t}a_{\text{at},i}^{(n)\dagger}(t)&=& -i\omega_{\rm{at}}a_{\text{at},i}^{(n)\dagger}(t)+i\sum_{m}c_{i}^{(n,m)\dagger}(t),\label{eq:heisenberg-atom}\\
&&+i\Omega_{\text{R}}a_{i}(t)\nonumber\\
\partial_{t}b_{i}^{(m)}(t)&=&-i\omega_{m}b_{i}^{(m)}(t)-i g_m a_{i}(t),\label{eq:heisenberg-loss}\\
\partial_{t}c_{i}^{(n,m)\dagger}(t)&=&-i\tilde{\omega}_{m}c_{i}^{(m)\dagger}(t)-i \tilde{g}_ma_{\text{at},i}^{(n)\dagger}(t).\label{eq:heisenberg-pump}
\end{eqnarray}

Injecting the integrated equation (\ref{eq:heisenberg-pump}) for the bath oscillators into the equation (\ref{eq:heisenberg-atom}) for the atomic degrees of freedom, we obtain a Markovian quantum langevin equation for the atomic field coupled to the photonic field :
\begin{equation}
\partial_{t}a_{\text{at},i}^{(n)\dagger}(t)=\left(-i\omega_{\rm{at}}^{(n)}-\frac{\Gamma_{\rm{p}}^{\rm{at}}}{2}\right)a_{\text{at},i}^{(n)\dagger}(t)
+ i\Omega_{\text{R}}a_{i}(t)+\hat{\xi}_{\text{at},i}^{(n)}(t)\label{eq:quantum-langevin-eq_atom_photon}
\end{equation}
with a Markovian quantum noise contribution related to atomic pumping :
\begin{eqnarray}
\left\langle \hat{\xi}_{\text{at},i}^{(n)\dagger}(t+\tau)\hat{\xi}_{\text{at},i}^{(n')}(t)\right\rangle &=&\delta_{i,j}\delta_{n,n'}\Gamma_{\rm{p}}^{\rm{at}}\delta{(\tau)},\\
\left\langle \hat{\xi}_{\text{at},i}^{(n)}(t+\tau)\hat{\xi}_{\text{at},i}^{(n')\dagger}(t)\right\rangle &=&0.
\end{eqnarray}
Then, integrating Eqs.~(\ref{eq:quantum-langevin-eq_atom_photon}),(\ref{eq:heisenberg-loss}) and injecting them in Eq.~(\ref{eq:heisenberg-photon}) we get for the photonic dynamics :
\begin{multline}
\label{eq:finite_time_langevin}
\partial_{t}a_{i}(t)=-i\com{a_{i}(t)}{H_{\rm{ph}}(t)}-\int_{t'} \Gamma_{\rm{l}}(t') a_i(t-t') 
+\hat{\xi}_{\text{l},i}(t)\\
+\int_0^t ds \left(\sum_{n}\Omega_{\text{R}}^2e^{(-i\omega_{\rm{at}}^{(n)}-\frac{\Gamma_{\rm{p}}^{\rm{at}}}{2})(t-s)}a_{i}(s)\right)\\
-i\Omega_{\text{R}}\int_0^t ds \sum_n e^{(-i\omega_{\rm{at}}^{(n)}-\frac{\Gamma_{\rm{p}}^{\rm{at}}}{2})(t-s)} \hat{\xi}^{(n)}_{\text{at},i}(s)\\
-i\Omega_{\text{R}}e^{(-i\tilde{\omega}_{m}-\frac{\Gamma_{\rm{p}}^{\rm{at}}}{2})t}\sum_{n} a_{\text{at},i}^{(n),\dagger}(0),
\end{multline}
where the expressions for the loss memory kernel and noise autocorrelations are described below. 

\subsection{Langevin equation: general form}
At long times with respect to $1/\Gamma_{\rm{p}}^{\rm{at}}$, the time-dependent contribution $\propto e^{(-i\tilde{\omega}_{m}-\Gamma_{\rm{p}}^{\rm{at}}/2)t}\sum_{n} a_{\text{at},i}^{(n)\dagger}(0)$ in Eq.~(\ref{eq:finite_time_langevin}) (which represents a memory of the initial conditions) vanishes, and we can also replace the boundaries in the various integrals by $0$ and $+\infty$. We obtain then the final form for the photonic non-Markovian Langevin equation of Eq.~(\ref{eq:quantum-langevin-Bose})
\begin{multline}
\partial_{t}\hat{a}_{i}(t)=-i\com{\hat{a}_{i}(t)}{H_{\rm{ph}}(t)}\\
+\int_{-\infty}^\infty d\tau [\Gamma_{\rm{p}}(\tau)-\Gamma_{\rm{l}}(\tau)]\hat{a}_{i}(t-\tau)+\hat{\xi}_{\text{p},i}(t)+\hat{\xi}_{\text{l},i}(t)\label{eq:quantum-langevin-eq}
\end{multline} 
where
\begin{equation}
\hat{\xi}_{\text{p},i}(t)= -i\Omega_{\text{R}}\int_{-\infty}^t ds \sum_n e^{(-i\omega_{\rm{at}}^{(n)}-\frac{\Gamma_{\rm{p}}^{\rm{at}}}{2})(t-s)} \hat{\xi}^{(n)}_{\text{at},i}(s).
\end{equation}
The non-zero contributions for the two-points quantum noise autocorrelations can be summarized into :
\begin{equation}
\begin{array}{lll}
\left\langle \hat{\xi}_{\text{l},i}(t+\tau)\hat{\xi}^{\dagger}_{\text{l},j}(t)\right\rangle &=&\delta_{i,j}\int_{\omega}\mathcal{S}_{\rm{l}}(\omega)e^{-i\omega\tau}\\
\left\langle \hat{\xi}_{\text{p},i}^{\dagger}(t+\tau)\hat{\xi}_{\text{p},j}(t)\right\rangle &=&\delta_{i,j}\int_{\omega}\mathcal{S}_{\rm{p}}(\omega)e^{+i\omega\tau}
\end{array}
\label{eq:noise-autocorrelation}
\end{equation}
where $\Gamma_{\rm{l}}(\tau)=\theta(\tau)\int_{\omega}\mathcal{S}_{\rm{l}}(\omega)e^{-i\omega\tau}$ and $\Gamma_{\rm{p}}(\tau)=\theta(\tau)\int_{\omega}\mathcal{S}_{\rm{p}}(\omega)e^{-i\omega\tau}$. While the loss power spectrum  $\mathcal{S}_{\rm{l}}(\omega)$ is provided in Eq.~(\ref{eq:phys-system-bath-loss-spectral-density}), the photonic pump power spectrum has the expression
\begin{equation}
\label{eq:derivation_emission_power spectrum}
\mathcal{S}_{\rm{p}}(\omega)=\Gamma_{\text{p}}^{(1)}\int d\omega' \mathcal{D}(\omega')\frac{(\Gamma_{\rm{p}}^{\rm{at}}/2)^2}{(\omega-\omega')^2+(\Gamma_{\rm{p}}^{\rm{at}}/2)^2},
\end{equation}
where $\Gamma_{\text{p}}^{(1)}=4\Omega_{\text{R}}^2/\Gamma_{\rm{p}}^{\rm{at}}$ is the maximum photonic pumping rate for a single atom, and is obtained at resonance:  as in \cite{Lebreuilly_2016,Lebreuilly_square}, each atom is responsible for a Lorentzian contribution to the photonic pumping, the continuous sum of the various contributions then provides the full spectrum $\mathcal{S}_{\rm{p}}(\omega)$.
\subsection{Some examples of realizable power spectra}
\subsubsection{First example: Markovian losses and Lorentzian pump power spectra}\label{subsec:engineer-lorentzian-spectrum}
As a first example, we set ourselves in the configuration in which losses are Markovian processes, i.e., $\mathcal{S}_{\rm{p}}(\omega)=\Gamma_{\rm{l}}$, and all atomic transitions are equal to $\omega_{\rm{p}}$, in such a way that $\mathcal{D}(\omega)=N_{\rm{at}}\delta(\omega-\omega_{\rm{p}})$. In that case we obtain for the photonic pump power spectrum the Lorenzian form:
\begin{equation}
\label{eq:derivation_emission_power spectrum_lorenzian}
\mathcal{S}_{\rm{p}}(\omega)=N_{\rm{at}}\Gamma_{\text{p}}^{(1)} \frac{(\Delta_{\text{diss}}/2)^2}{(\omega-\omega_{\rm{p}})^2+(\Delta_{\text{diss}}/2)^2},
\end{equation}
where we have set the value $\Gamma_{\rm{p}}^{\rm{at}}=\Delta_{\text{diss}}$ for the atomic pumping rate. This configuration leads to the specific model Eq.~(\ref{eq:lorentzian-Markovian-power-spectra}) introduced in Sec.~\ref{sec:quantum-Langevin} that we have chosen in order to perform numerical simulations.
\subsubsection{Second example: artificial Kennard-Stepanov relation}\label{subsec:artificial-Kennard-Stepanov}
Another option would be to engineer non-trivial distributions $\mathcal{D}(\omega)$ (which we could imagine to do, e.g., by tuning all atoms to different frequencies, or by using several atomic species) of the atomic transition frequencies in such a way to simulate a Kennard-Stepanov relation. More specifically, we choose losses to be also Markovian $\mathcal{S}_{\rm{l}}(\omega)=\Gamma_{\rm{l}}$, and the particular form
\begin{equation}
\label{eq:exponential-distribution}
\mathcal{D}(\omega)=\mathcal{D}_0 e^{\beta_{\text{eff}}\omega}.
\end{equation}
 for the distribution of atomic transition frequencies. In that case the pump power spectrum becomes
\begin{equation}
\label{eq:derivation_emission_power spectrum_KS_approx}
\mathcal{S}_{\rm{p}}(\omega)=\mathcal{D}_0 \Gamma_{\text{p}}^{(1)}\int d\omega' e^{\beta_{\text{eff}}\omega'}\frac{(\Gamma_{\rm{p}}^{\rm{at}}/2)^2}{(\omega-\omega')^2+(\Gamma_{\rm{p}}^{\rm{at}}/2)^2}.
\end{equation}
In the limit of a very weak atomic pumping rate $\Gamma_{\rm{p}}^{\rm{at}}\ll T_{\text{eff}}=1/\beta_{\text{eff}}$, we recover the exponential-shaped spectrum:
\begin{equation}
\label{eq:derivation_emission_power spectrum_KS}
\mathcal{S}_{\rm{p}}(\omega)=\Gamma_{\text{p}}e^{\beta_{\text{eff}}\omega},
\end{equation}
where $\Gamma_{\text{p}}=\frac{\pi}{2} \mathcal{D}_0\Delta_{\text{diss}} \Gamma_{\text{p}}^{\rm{at}}$. The Kennard-Stepanov relation Eq.~(\ref{eq:exact-KS}) is thus reproduced artificially even though the photonic environment is highly out-of-equilibrium. Theoretically, this spectral shape (initially proposed in \cite{Shabani_artificial_thermal_bath}) can be reproduced  for an arbitrary temperature: if necessary one can lower simultaneously the pumping rate $\Gamma_{\rm{p}}^{\rm{at}}$ and $\Omega_{\text{R}}$, while increasing the number of atoms in order to stay within the previously described conditions of validity of the quantum Langevin equation (\ref{eq:quantum-langevin-eq}). Concretely, for very low $T_{\rm{eff}}$ the engineering procedure might be become more complex as it requires an high number of emitters with a fine control on transition frequencies.

\subsection{Pseudo-thermalization in exciton-polaritons and VSCSEL experiments}
\label{subsec:excitons}
The artificial Kennard-Stepanov configuration mentioned in Sec.~\ref{subsec:artificial-Kennard-Stepanov} might also be naturally reproduced in low-T exciton-polaritons experiments \cite{Kasprzak_BEC,Balili_BEC}.

While most theoretical works in the early literature \cite{Porras_scattering,Malpuech_exciton_scattering} have stressed on the impact of exciton-exciton scattering processes in the relaxation of polaritons into the bottleneck region of lower branch, recent works \cite{Maragkou_optical_phonons} have raised the possibility that high energy longitudinal optical (LO) phonons might play an important role in the polariton relaxation dynamics in some regimes. We discuss here what might be the implications regarding the nature of thermalization in such physical situation.

Since the excitons (located in an higher energy with respect to the bottom of the polaritonic band) usually undergo fast collisions/energy exchanges processes and also possess a much longer lifetime than polaritons, the exciton reservoir is rather well thermalized (while polaritons might not be able to thermalize among them) and can thus be described by a classical Boltzmann distribution $n_{X}(\epsilon^X_{k})\propto e^{-\beta \epsilon^X_{k}}$ (excitons being very massive particles, their degree of degeneracy is usually very weak in those experiments). 

One hand, since the LO phonons dispersion law is typically very flat and strongly located around the frequency $\omega_{\rm{LO}}$ (in stark contrast with acoustic phonons whose dispersion law present a light-cone structure), the LO phonon-assisted scattering processes excitons$\to$polaritons maintain the full information on the excitonic energy distribution and transfer it into the frequency-dependence of the polariton injection rate (up to an energy shift  $\hbar \omega_{\rm{LO}}$): in the hypothesis that LO phonon-assisted scattering processes are dominant, the polaritonic injection rate should thus present an exponential frequency dependence ($\mathcal{S}_{\rm{p}}(\omega)\simeq \Gamma_{\rm{p}}^{P} e^{-\beta_{\rm{eff}}\omega}$) at a good degree of approximation. 

On the other hand, in that same picture, polariton$\to$exciton recombination processes are strongly inhibited as they would involve the absorption of a phonon from the LO phononic reservoir, which can be approximated as being close to the vacuum state (LO phonons possessing a significantly higher energy ($\simeq 5\, meV$) than the typical temperatures ($\simeq 0.5\, meV$) in exciton-polaritons). As a consequence, polaritonic losses are by far dominated by mirror transparency effects, and can be well represented by Markovian processes: $\mathcal{S}_{\rm{l}}(\omega)\simeq \Gamma^{\rm{P}}_{\rm{l}}=x_{\rm{ph}}\Gamma_{\rm{ph}}$, where $x_{\rm{ph}}$ is the photonic fraction in the bottom of the lower polaritonic branch, and $\Gamma_{\rm{ph}}$ is the photonic loss rate.  One concludes that the Kennard-Stepanov relation $\mathcal{S}_{\rm{p}}(\omega)/\mathcal{S}_{\rm{l}}(\omega)\simeq \Gamma_{\rm{p}}^{\rm{P}}/\Gamma_{\rm{l}}^{P}  e^{-\beta_{\rm{eff}}\omega}$ might be artificially verified in that context (at least in a broad frequency region), and polaritons be subject to pseudo-thermalization. 

Even more importantly, a similar phenomenology may be invoked to explain the peculiar features observed in the VCSEL device of~\cite{Bajoni_BEC}: as the excitonic-polaritonic strong coupling is broken by the high density of excitations present in the active medium, scattering between bare photons is expected to be very inefficient. The observed thermal distribution of photons can therefore be hardly explained in terms of standard collisional thermalization within the gas of photons, but must be inherited by energy exchange processes with the external environment, which can be well represented by the combination of an amplifying reservoir formed of thermalized free carriers and a dissipative reservoir due transparency of the cavity mirrors. Here again, the Kennard-Stepanov relation might be artificially verified in specific configurations where the processes of absorption by free carriers are inhibited and thus weak with respect to the rate of particle losses, leading to an apparent photonic thermalization.

{Based on these arguments, the measurement of thermal signatures in the polaritonic (resp. photonic) observables in exciton-polariton (resp. VCSEL) experiments has to be interpreted carefully. On one hand, one should first experimentally investigate whether the thermal-like momentum distribution is associated to a satisfied FDT using, e.g., the protocol proposed in~\cite{Chiocchetta_thermalization}. Then, before drawing any conclusion regarding a true thermalization or a pseudo-thermalization, one should also verify that polaritons (resp. photons) are indeed equilibrated with their environment of excitons (resp. free carriers), and phonons: to this purpose,} one way to proceed would be to check the validity of the FDT associated to a pair of operators $\hat{A}(t)$ and $\hat{B}(t)$ (with the notations of Sec.~\ref{sec:FDT}) associated respectively to polariton and the reservoirs degrees of freedoms, by measuring the corresponding frequency-dependent effective temperature.
\section{How to break pseudo-thermalization}
\label{sec:out-of-equilibrium}
Expectedly, the low-energy pseudo-thermalization effect described in Sec.~\ref{sec:effective-thermalization} is not a fully general properties of driven-dissipative quantum systems, since a wide class of models can not been cast into the form of the quantum Langevin Eq.~(\ref{eq:quantum-langevin-Bose}), which only implements non-Markovian loss and pump processes, and does not include many other possible effects such as the saturation of the emitters or dephasing. 

In this section, we discuss a simple extension of Eq.~(\ref{eq:quantum-langevin-Bose}) which allows to break the emergent equilibrium presented in Sec.~\ref{sec:effective-thermalization}. More specifically, we introduce a generalized Bogoliubov-de Gennes model at low energies and low momenta, with a complex kinetic energy and a complex chemical potential:
\begin{multline}
\label{eq:out-of-eq-linear-langevin}
-i\omega\hat{\Lambda}_\kk (\omega)=-i\left[z\epsilon_\kk \hat{\Lambda}_\kk (\omega)+\tilde{z}\mu\right.\\
\left.\left( \hat{\Lambda}_\kk (\omega)+\hat{\Lambda}_{-\kk}^\dagger (-\omega)\right)\right]+\hat{\xi}_{neq,\kk}(\omega).
\end{multline}
The noise auto correlation is
\begin{align}
\label{eq:out-of-eq-noise}
\langle \hat{\xi}_{neq,\kk}(\omega)\hat{\xi}^{\dagger}_{neq,\kk '}(\omega')\rangle  &= \langle \hat{\xi}^\dagger_{neq,\kk}(\omega)\hat{\xi}_{neq,\kk }(\omega')\rangle\\
&=\delta_{\kk-\kk'}\, \delta_{\omega-\omega'}\mathcal{S}_{\rm{l}}(\omega_{\text{BEC}}),\nonumber
\end{align}
and complex couplings are written in phase-modulus representation as $z=\rho\ee^{-i\theta}$, $\tilde{z}=\tilde{\rho}\ee^{-i\tilde{\theta}}.$ 
This model is very similar to the low-energy model Eq.~(\ref{eq:low-energy-langevin-frequency}) derived in a previous section, except that the kinetic energy $\epsilon_k$ and the chemical potential $\mu$ have respectively been multiplied by two different complex numbers $z$ and $\tilde{z}$ (while they were multiplied by the same complex in the low-energy theory Eq.~(\ref{eq:low-energy-langevin-frequency}). In Sec.~\ref{sec:static-out-of-eq} and Sec.~\ref{sec:FDT-out-of-eq} we will show that in case of alignement in the complex plane of these couplings (i.e., $\theta=\tilde{\theta}$), we obtain an effective equilibrium theory, while in the case of a misalignement, the steady state presents non-equilibrium features. Finally in Sec.~\ref{sec:driving-out-of-of-equilibrium-examples}, we will describe a few ways to implement those modified complex couplings.

\subsection{Static correlations}
\label{sec:static-out-of-eq}
Analysing Eqs.~(\ref{eq:out-of-eq-linear-langevin}),(\ref{eq:out-of-eq-noise}) we obtain the following expression for the static  momentum distribution $n^{neq}_{k}=\langle\hat{\Lambda}^\dagger_\kk\hat{\Lambda}_{\kk}\rangle $ and the anomaleous average $\mathcal{A}^{neq}_{k}=\langle\hat{\Lambda}_\kk\hat{\Lambda}_{-\kk}\rangle $ (the derivation is very similar to the one made in App.~\ref{app:static-correlations}):
\begin{small}
\begin{equation}
\label{eq:non-eq-static-correlations}
n^{neq}_{k}=\frac{|z\epsilon_k+\tilde{z}\mu|^2 \mathcal{S}_{\rm{l}}(\omega_{\text{BEC}})/2}{\left(\rho \mathrm{sin}(\theta)\epsilon_k+\tilde{\rho}\mathrm{sin}(\tilde{\theta})\mu\right)\rho \epsilon_k \left(\rho\epsilon_k+2\mathrm{cos}(\theta-\tilde{\theta})\tilde{\rho}\mu\right)},
\end{equation}
\end{small}
\begin{small}
\begin{equation}
\label{eq:non-eq-static-correlations-anomaleous}
\mathcal{A}^{neq}_{k}=\frac{-(z^*\epsilon_k+\tilde{z}^*\mu)\tilde{z}\mu  \mathcal{S}_{\rm{l}}(\omega_{\text{BEC}})/2}{\left(\rho \mathrm{sin}(\theta)\epsilon_k+\tilde{\rho}\mathrm{sin}(\tilde{\theta})\mu\right)\rho \epsilon_k \left(\rho\epsilon_k+2\mathrm{cos}(\theta-\tilde{\theta})\tilde{\rho}\mu\right)}.
\end{equation}
\end{small}

In the general case, it is not possible to further simplify those expressions, and the steady-state properties differs from the equilibrium statistics, as static correlations can not be cast in the form of a Rayleigh-Jeans thermal law (e.g., for the momentum distribution $n_{k}=T_{\rm{eff}}(\epsilon_k+\mu)/E_{k}^2$). However, considering the particular case in which the complex couplings $z$ and $\tilde{z}$ are aligned in the complex plane, i.e., $\theta=\tilde{\theta}$, one obtains
\begin{eqnarray}
\label{eq:non-eq-static-correlations-aligned}
n^{aligned}_{k}&=&\frac{\tilde{T}_{\text{eff}}(\rho\epsilon_k +\tilde{\rho}\mu)}{\rho\epsilon_k (\rho\epsilon_k +2\tilde{\rho}\mu)},\\
\label{eq:non-eq-static-correlations-anomaleous-aligned}
\mathcal{A}^{aligned}_{k}&=&\frac{-\tilde{T}_{\text{eff}}\tilde{\rho}\mu}{\rho\epsilon_k (\rho\epsilon_k +2\tilde{\rho}\mu)},
\end{eqnarray}
which compared to Eqs.~(\ref{eq:thermal-static-correlations}),~(\ref{eq:thermal-static-correlations-anomaleous}), corresponds to a low-energy effective equilibrium statistics with \begin{equation}
\tilde{T}_{\text{eff}}=\frac{\mathcal{S}_{\rm{l}}(\omega_{\text{BEC}})}{2\text{sin}(\theta)}                                                                                                                                                                                                                                                                                                                                           \end{equation}
and renormalized couplings $\epsilon_k\to \rho\epsilon_k$, $\mu\to \tilde{\rho}\mu$. This is not surprising since in that case, the generalized Bogoliubov-de Gennes model given by Eq.~(\ref{eq:out-of-eq-linear-langevin}) coincides with the low frequency limit Eq.~(\ref{eq:low-energy-langevin-frequency}) of the non-Markovian Langevin equation studied in this paper. We conclude that the alignement configuration of the couplings  $z$ and $\tilde{z}$ of Eq.~(\ref{eq:low-energy-langevin-frequency}) corresponds to an effective equilibrium situation, while the general case of non-alignement drives the system out-of-equilibrium, as thoroughly discussed in \cite{Sieberer_3D,Sieberer_BEC,Altman_2D}.

Although Eqs.~(\ref{eq:non-eq-static-correlations}),~(\ref{eq:non-eq-static-correlations-anomaleous}) present deviations from the Rayleigh-Jeans law for $E_k\to 0$, for a generic choice of misalignement of $z$ and $\tilde{z}$ the low-momentum correlations still present a $1/k^2$ equilibrium-like infrared divergence and we do not expect any particular loss of coherence by driving the system out-of-equilibrium,  at least in three or higher dimensions. This is generically valid except for the specific pathological configuration in which we set the phase {\small $\theta$} to $0$ and the phase {\small $\tilde{\theta}$} to $\pi/2$. In this case, which can be obtained by using Markovian baths, cancelling the photon-photon interactions and adding saturation to the pump (see Sec.~\ref{sec:saturation-emitter}), we indeed obtain a very different behaviour
\begin{eqnarray}
\label{eq:non-eq-static-correlations-critic}
n^{pathological}_{k}&=&\frac{\mathcal{S}_{\rm{l}}(\omega_{\text{BEC}})(\epsilon_k^2+(\tilde{\rho}\mu)^2)}{2\tilde{\rho}\mu\epsilon_k^2},\\
\label{eq:non-eq-static-anomaleous-correlations-critic}
\mathcal{A}^{pathological}_{k}&=&\frac{i \mathcal{S}_{\rm{l}}(\omega_{\text{BEC}})(\epsilon_k+i\tilde{\rho}\mu)\tilde{\rho}\mu}{2\tilde{\rho}\mu\epsilon_k^2}.
\end{eqnarray}
We see that the momentum distribution changes behaviour at long range : $n(k)\simeq1/k^4$, such a feature has already been predicted in \cite{Chiocchetta_Langevin}. 

Due to these increased low-momenta fluctuations, we might be tempted to conclude that in three dimensions, a non-equilibrium free Bose gas in presence of a pump and saturation, i.e., a 3D VCSEL \cite{VCSEL} can not Bose-condense (while the equilibrium free Bose gas is known to condense). However, in this case the Bogoliubov approach is inconsistent and can not be applied in a straightforward manner since the nonlinear corrections are very large for small $k$ modes and can not be neglected. 

Instead, accessing the long range properties in this regime requires applying the renormalization group methods to this non-equilibrium system while keeping all relevant non-linearities (including those providing from saturation effects): our understanding is that during the RG flow \cite{Sieberer_3D,Sieberer_BEC}, a small photon-photon interaction should be generated  and the true correlations should be thus in $n(k)\simeq 1/k^2$, saving thus the convergence. Such effect was verified numerically in \cite{Ji_temporal_1D} by simulating the Kardar-Parisi-Zhang equation (however in that case the simulations were done in a 1D configuration).

\subsection{Momentum-dependent effective temperatures from the FDT}
\label{sec:FDT-out-of-eq}
\begin{figure}[t]
\centering
\includegraphics[height=1.4\columnwidth,clip]{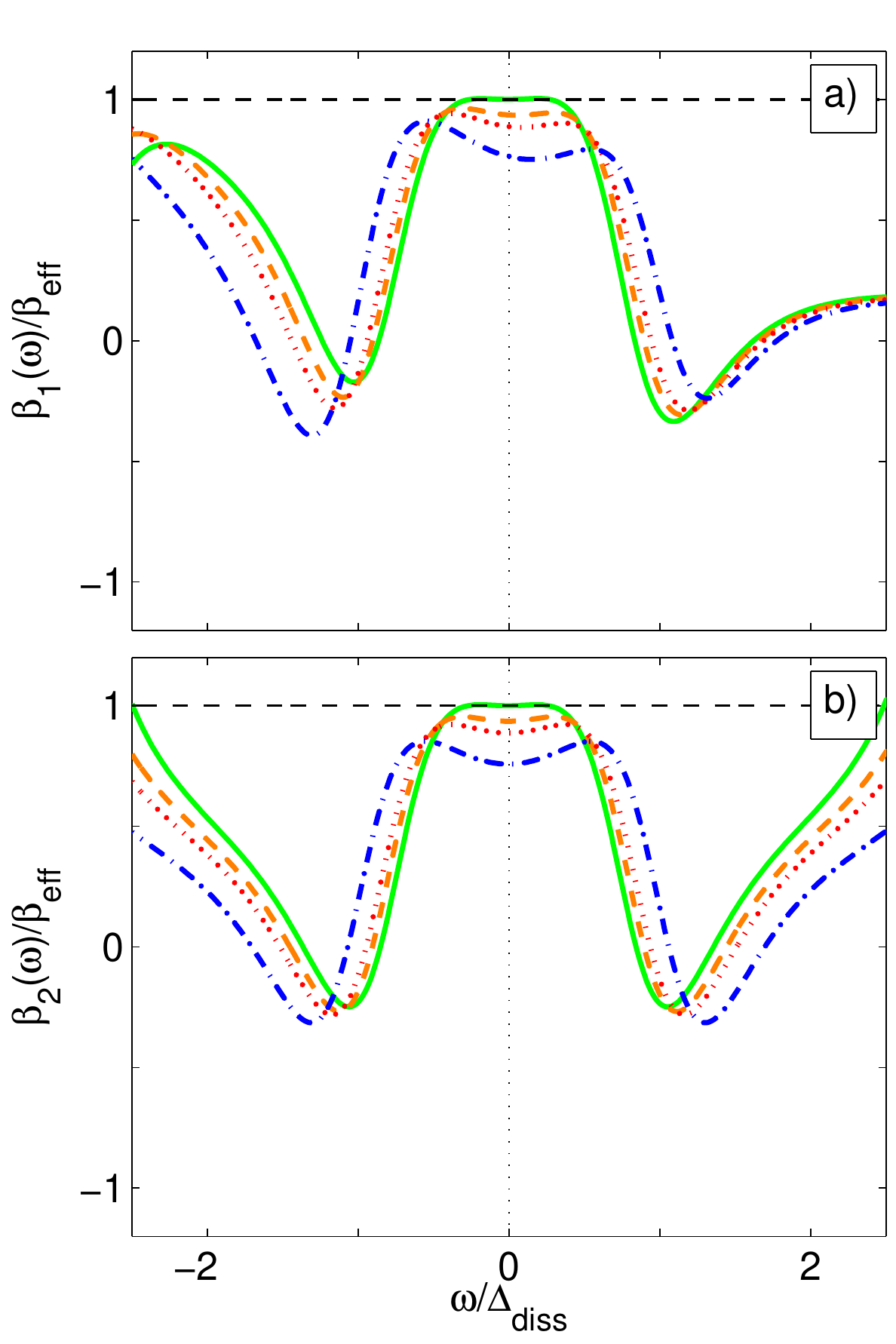}
\caption{\label{fig:FDT-dispersion}Test of the FDT/KMS relation in the presence of dispersion of the emitters. Panel a) (resp. b)): frequency-dependent effective temperature $\beta_{1}(\kk,\omega)$ (resp. $\beta_{2}(\kk,\omega)$) defined in Eq.~(\ref{eq:effective-temperature}) for a Lorentzian pump, mobile and massive emitters and Markovian losses (model defined in Sec.\ref{sec:driving-out-of-of-equilibrium-dispersion}), in function of the frequency $\omega$ in units of $\Delta_{\text{diss}}$, and for various momenta $\kk$. Parameters:  $M_{\rm{p}}/m=3$, $\Gamma_{l}/\Gamma_{\rm{p}}^0=0.3$, $\Gamma_{\rm{p}}/\Delta_{\text{diss}}=0.01$, $\delta/\Delta_{\text{diss}}=-2$. For each panel, the various curves
correspond to increasing values of the momentum $k$, chosen in such a way that the corresponding Bogoliubov energies span a wide energy range across the effective temperature $T^{\rm{disp}}_{\rm{eff}}\equiv 1/\beta_{\rm{eff}}^{\rm{disp}}=0.54\Delta_{\rm{diss}}$): $k/k_{\rm{th}}=0.18$ for the green solid line, $k/k_{\rm{th}}=3.65$, for the orange dashed line, $k/k_{\rm{th}}=9.1$ for the red dotted line, $k/k_{\rm{th}}=54.7$ for the dash-dotted blue line: $k/k_{\rm{th}}=3\times 10^{-2}$ for the green solid line, $k/k_{\rm{th}}=1.83$, for the orange dashed line, $k/k_{\rm{th}}=2.43$ for the red dotted line, $k/k_{\rm{th}}=3.66$ for the  dash-dotted blue line. Here  $k_{\rm{th}}$ is also defined by $E(\kk_{\text{th}})= T^{\rm{disp}}_{\rm{eff}}$}
\end{figure}
It is also interesting to check whether a misalignement of the couplings affects the validity of the FDT. To do so, we will use an exact model providing a quantum Langevin equation valid at all frequencies which leads at low-frequencies and low-momenta to the effective description Eq.~(\ref{eq:out-of-eq-linear-langevin}) with non-aligned couplings $z$ and $\tilde{z}$: this model is defined in the next section in Eqs~(\ref{eq:dispersion-quantum-langevin-momentum}),~(\ref{eq:power-spectrum-dispersion}). We computed the corresponding effective temperatures $\beta_{1}(\kk,\omega)$ and $\beta_{2}(\kk,\omega)$ by mean of the definitions  Eqs~(\ref{eq:effective-temperatures}). 

In Fig.~\ref{fig:FDT-dispersion}, we show $\beta_{1}(\kk,\omega)$ (resp. $\beta_{2}(\kk,\omega)$) in the left panel (resp. right panel) in function $\omega$ in units of $\Delta_{\text{diss}}$ for various momenta $\kk$: we notice that in the region $|\omega|\ll \Delta_{\text{diss}}$, these effective temperatures do not take anymore identical values, but indeed tend toward a momentum-dependent value. We conclude that pseudo-thermalization is broken not only at a static level (in the sense that it does not respect perfectly the Rayleigh-Jeans law obtained for a weakly interacting isolated Bose gas) in case of misalignement, but also at a dynamical level, as the FDT is not verified at low-frequencies.
\subsection{Examples of modified quantum optics models driving the system out-of-equilibrium}
\label{sec:driving-out-of-of-equilibrium-examples}
In this section, we discuss various physical ways to obtain the modified Bogogliubov-de Gennes system Eq.~(\ref{eq:out-of-eq-linear-langevin}) with misalignement of the complex couplings, by mean of simple modifications with respect to the quantum optics model introduced in Sec.~\ref{sec:Langevin-derivation}. 
\subsubsection{Emitters with dispersion}
\label{sec:driving-out-of-of-equilibrium-dispersion}
The first model we introduce is very similar to the one presented in Sec.~\ref{sec:quantum-Langevin}, except that we add a momentum-dependence to the photonic pump power spectrum $S_{\text{p},\kk}(\omega)$. In the quantum optics model presented in Sec.~\ref{sec:Langevin-derivation}, this can be obtained  by taking into account the recoil of the mobile and massive two-level atoms which leads to the expression:
\begin{equation}
\label{eq:power-spectrum-dispersion}
S_{\text{p},\kk}(\omega)=\Gamma_{\rm{p}}\frac{(\Delta_{\text{diss}}/2)^2}{(\omega+\epsilon_\kk^{\text{p}}-\omega_{\rm{p}})^2+(\Delta_{\text{diss}}/2)^2}
\end{equation} 
with $\epsilon_\kk^{\text{p}}=k^2/2M_{\rm p}$ defined as the recoil energy for the emission of a photon of a wave vector $k$.
If the mass $M_{\rm{p}}$ of the emitters is small enough, this effect can be physically relevant. We obtain thus the following Langevin equation:
\begin{multline}
\label{eq:dispersion-quantum-langevin-momentum}
\frac{\partial}{\partial t}\hat{\psi}_{\kk}(t)=-i\com{\hat{\psi}_{\kk}(t)}{H_{\rm{ph}}(t)}\\
+\int_{-\infty}^\infty d\tau [\Gamma_{p,\kk}(\tau)-\Gamma_{\rm{l}}(\tau)]\hat{\psi}_{\kk}(t-\tau)
+\hat{\xi}_{disp,\kk}(t),
\end{multline}
with the non-Markovian momentum-dependent dissipative kernel for pumping
\begin{equation}
\label{eq:dispersion-emission-kernel}
\Gamma_{\text{p},\kk}(\tau)=\Theta(\tau)\int_{\omega}S_{\text{p},\kk}(\omega)\ee^{-i\omega \tau},
\end{equation}
and noise correlations in momentum-frequency space
\begin{subequations}
\label{eq:noise-correlation-dispersion-frequency-momentum}
\begin{align}
\langle \hat{\xi}_{disp,\kk}(\omega)\hat{\xi}^{\dagger}_{disp,\kk '}(\omega')\rangle & = \delta_{\kk-\kk'}\, \delta_{\omega-\omega'}\mathcal{S}_{\rm{l}}(\omega_{\text{BEC}}+\omega), \\
\langle \hat{\xi}^{\dagger}_{disp,\kk}(\omega)\hat{\xi}_{disp,\kk'}(\omega')\rangle & =\l\delta_{\kk-\kk'}\, \delta_{\omega-\omega'} S_{\text{p},\kk}(\omega_{\text{BEC}}+\omega).
\end{align}
\end{subequations}
We used the theory Eq.~(\ref{eq:dispersion-quantum-langevin-momentum}), and applied the Bogoliubov methods in order to compute analytically the correlation functions in momentum-frequency space. In order to test the FDT, we define for this specific model the physical quantity 
\begin{equation}
\beta_\text{eff}\equiv = \frac{\dd }{\dd \omega}\log\left[\frac{\mathcal{S}_{\rm{l}}(\omega)}{\mathcal{S}_{\text{em},k}(\omega)}\right]\biggr|_{\omega = \omega_{\rm{BEC}},k=0}
\end{equation}
which we plotted in dashed horizontal lines in  Fig.~\ref{fig:FDT-dispersion}a)-b) and compared to the momentum-frequency dependent inverse temperatures $\beta_1(\kk,\omega)$ and $\beta_2(\kk,\omega)$ of Eq.~(\ref{eq:effective-temperatures}). 


Still in the Bogoliubov regime, from Eq.~(\ref{eq:dispersion-quantum-langevin-momentum}) we can derive a low-energy and low-momentum effective theory by applying a procedure similar to Sec.\ref{sec:low-frequency-dynamics}:
\begin{multline}
\label{eq:out-of-eq-linear-langevin-dispersion}
-i\omega\hat{\Lambda}_\kk (\omega)=-i\left[z_{disp}\epsilon_\kk \hat{\Lambda}_\kk (\omega)\right.\\
\left.+\tilde{z}_{disp}\mu\left( \hat{\Lambda}_\kk (\omega)+\hat{\Lambda}_{-\kk}^\dagger (-\omega)\right)\right]+\bar{\xi}_{disp,\kk}(\omega).
\end{multline}
The noise correlations are
\begin{subequations}
\begin{align}
\langle \bar{\xi}_{disp,\kk}(\omega)\bar{\xi}^{\dagger}_{disp,\kk'}(\omega')\rangle & = \delta_{\kk-\kk'}\, \delta_{\omega-\omega'}\mathcal{S}_{\rm{l}}(\omega_{\text{BEC}}), \\
\langle \bar{\xi}^{\dagger}_{disp,\kk}(\omega)\bar{\xi}_{disp,\kk'}(\omega')\rangle & =\l\delta_{\kk-\kk'}\, \delta_{\omega-\omega'} S_{\text{p},0}(\omega_{\text{BEC}}),
\end{align}
\end{subequations}
where $S_{\text{p},0}(\omega_{\text{BEC}})=\mathcal{S}_{\rm{l}}(\omega_{\text{BEC}})$ and the complex couplings are
\begin{align}
z_{disp}&=(1+\tilde{\delta}-i\tilde{\Gamma})(1+i2M_{\rm{p}}\underbrace{\partial^2_k\,\Gamma_{\rm{p}}\,_{|k=0,\omega=\omega_{\text{BEC}}}}_{<0}),\\
\tilde{z}_{disp}&=(1+\tilde{\delta}-i\tilde{\Gamma}).
\end{align}
We obtain some effective complex kinetic energy and chemical potential for the photonic dynamic. However, as predicted, due to the dispersion of the emitters an additional multiplicative contribution has been added to the complex kinetic energy inducing thus a phase misalignement between  the complex terms $z_{disp}$ and $\tilde{z}_{disp}$. 
\subsubsection{Saturation of the pump/two-body losses}
\label{sec:saturation-emitter}
In the second model, we propose to add saturation to the pump or two-body losses. Basing ourselves on the photonic case presented in Sec.~\ref{sec:Langevin-derivation}, some saturation can stem from the fact that the emitters are two-level atoms and thus are not perfectly linear systems. In this case, at a qualitative level the Langevin equation for the quantum fluctuations becomes at low frequency:
\begin{multline}
\label{eq:out-of-eq-linear-langevin-saturation}
-i\omega\hat{\Lambda}_\kk (\omega)=-i\left[z_{sat}\epsilon_k \hat{\Lambda}_\kk (\omega)\right.\\
\left.+\tilde{z}_{sat}\mu\left( \hat{\Lambda}_\kk (\omega)+\hat{\Lambda}_{-\kk}^\dagger (-\omega)\right)\right]+\bar{\xi}_{sat,\kk}(\omega).
\end{multline}
and the complex couplings are
\begin{align}
z_{sat}&=(1+\tilde{\delta}-i\tilde{\Gamma}),\\
\tilde{z}_{sat}&=(1+\tilde{\delta}-i\tilde{\Gamma})(1-i\gamma_{sat}).
\end{align}
$\gamma_{sat}$ is a dimensionless coupling quantifying the saturation effect, i.e, an increase of the dissipation strength with the density $\hat{\Lambda}_\kk^\dagger\hat{\Lambda}_\kk$, which linearized gives in the Bogoliubov approach a complex contribution proportional to $\hat{\Lambda}_\kk+\hat{\Lambda}_{-\kk}^\dagger$. Here again, because of saturation which multiplies the chemical potential by some complex, we also observe a misalignement between $z_{sat}$ and $\tilde{z}_{sat}$. 

For the sake of simplicity we assumed autocorrelations to be Gaussian at a first level of description:
\begin{subequations}
\label{Eq:noise_saturation}
\begin{align}
\langle \bar{\xi}_{sat,\kk}(\omega)\bar{\xi}^{\dagger}_{sat,\kk'}(\omega')\rangle & = \delta_{\kk-\kk'}\, \delta_{\omega-\omega'}\mathcal{S}_{\rm{l}}(\omega_{\text{BEC}}), \\
\langle \bar{\xi}^{\dagger}_{sat,\kk}(\omega)\bar{\xi}_{sat,\kk'}(\omega')\rangle & =\l\delta_{\kk-\kk'}\, \delta_{\omega-\omega'} S_{l}(\omega_{\text{BEC}}).
\end{align}
\end{subequations}
Yet, it is worth highlighting that, in presence of saturation, the noise should present non-trivial non-linear autocorrelations depending on the quantum field $\hat{\Lambda}_{\kk}$. Studying the effect of these corrections to gaussianity in the context of pseudo-thermalization will be the subject of a future work. The identity between both right-hand sides in Eq.~(\ref{Eq:noise_saturation}), which leads to an effective classical noise, is a consequence of the restriction to the regime low-momenta and low-frequencies, where a large average occupancy of each momentum state is expected above the BEC threshold, and non-classical effects related to the discrete nature of particles are rather weak corrections.

\section{Conclusion and perspectives}
\label{sec:conclusions}
In this work we have analysed the {\em pseudo-thermalization} effect, where an open quantum system coupled to several non-thermal and non-Markovian reservoirs presents an emergent thermal behavior in spite of the highly non-thermal nature of its environment. Our approach was based on  a quantum Langevin formalism which allows us to overcome the inherent issues related to the quantum master equation formalism and the quantum regression theorem in a non-Markovian context, and then to compute arbitrary multiple time correlators. The focus was set on the exactly solvable case of a driven-dissipative weakly interacting Bose-Einstein Condensate.

In particular, we have shown that pseudo-thermalization not only occurs  at the static level but is also accompanied by the satisfaction of the fluctuation-dissipation theorem at the dynamical level. According to the spectral properties of the chosen reservoirs, equilibrium signatures can be observed either only at low energies or globally. In the latter situation, which might relevant in some exciton-polariton and VCSEL experiments, the steady-state properties of the system alone are completely undistinguishable from the ones of an equilibrium system. Finally, several modifications of the initial model allowing to break this pseudo-thermalization effect have been discussed, with a particular stress on the role played by the dispersion and the saturation of the emitters.

The results of this work challenge the common idea that only open quantum systems in contact with an equilibrated environment can behave completely thermally. It implies in particular that,  before concluding to an equilibration, an experimentalist should check the thermal character not only of the system correlations, but also of the crossed correlations involving altogether the degrees of freedom of the system and the various reservoirs.

While this pseudo-thermalization effect is expected to be robust and universal with respect to the many-body dynamics of the considered physical system in the case where the Kennard-Stepanov relation is verified globally, it is unclear whether low-energy pseudo-thermalization should apply for any choice of system Hamiltonian in the generic case where the Kennard-Stepanov relation is only valid locally in frequency space: future studies will be dedicated in particular to the interplay between low-energy pseudo-thermalization and the departure of the Bogoliubov regime.

\acknowledgments The authors thank Emmanuele Dalla Torre for reading the manuscript and for his useful comments. Discussions with Maxime Richard, Andrea Gambassi, Laeticia Cugliandolo and Jamir Marino are also warmly acknowledged. JL and IC are supported by the EU-FET Proactive grant AQuS, Project
No. 640800, and by the Autonomous Province of Trento, partially through the project ``On silicon chip quantum optics for quantum computing and secure communications" (``SiQuro").   A.~C. acknowledges funding by the European Research Council via ERC Grant Agreement n. 647434 (DOQS). 

\appendix
\section{Quantum correlations in frequency and the FDT}\label{app:quantum-correlations}
In this Appendix, we compute the correlation matrix in momentum frequency space $\mathcal{C}_{\kk}(\omega)$ defined in Eq.~(\ref{eq:correlation-matrix}). We then move to the calculation of the momentum-frequency-dependent effective inverse temperatures involved in the test of the validity of the FDT, and defined in Eqs.~(\ref{eq:effective-temperatures}),(\ref{eq:effective-temperatures_anomaleous}). Inverting the Langevin equation in frequency space Eq.(\ref{eq:langevin-linear-system}), we get :
\begin{equation}
\left(\begin{array}{c}
\hat{\Lambda}_\kk(\omega)\\
\hat{\Lambda}^{\dagger}_{-\kk}(-\omega)
\end{array}\right)
=\frac{i}{\omega-\mathcal{L}_{\kk}(\omega)}
\left(\begin{array}{c}
\tilde{\xi}_{\kk}(\omega)\\
-\tilde{\xi}_{-\kk}^\dagger(-\omega)\end{array}\right)
\end{equation}
After calculation this gives us :
\begin{multline}
\left(\begin{array}{c}
\hat{\Lambda}_\kk(\omega)\\
\hat{\Lambda}^{\dagger}_{-\kk}(-\omega)
\end{array}\right)=\frac{i}{\begin{array}{l}\left[\omega-\left(\epsilon_{k}+\mu+i\tilde{\Gamma}(\omega)\right)\right]\\\times\left[\omega+\epsilon_{k}+\mu-i\tilde{\Gamma}^{*}(-\omega)\right]+\mu^2\end{array}}\\
\left(\begin{array}{c}
\left(\omega+\epsilon_{k}+\mu-i\tilde{\Gamma}^{*}(-\omega)\right)\tilde{\xi}_{\kk}(\omega)-\mu\tilde{\xi}^\dagger_{-\kk}(-\omega)\\
-\mu\tilde{\xi}_{\kk}(\omega)+\left(-\omega+\epsilon_{k}+\mu+i\tilde{\Gamma}(\omega)\right)\tilde{\xi}^\dagger_{-\kk}(-\omega)
\end{array}\right),
\end{multline}
and taking the hermitian conjugate:
\begin{multline}
\left(\begin{array}{c}
\hat{\Lambda}^\dagger_\kk(\omega)\\
\hat{\Lambda}_{-\kk}(-\omega)
\end{array}\right)=
\frac{-i}{\begin{array}{l}\left[\omega-\left(\epsilon_{k}+\mu-i\tilde{\Gamma}^{*}(\omega)\right)\right]\\
\times\left[\omega+\epsilon_{k}+\mu+i\tilde{\Gamma}(-\omega)\right]+\mu^2 \end{array}}\\
\left(\begin{array}{c}
\left(\omega+\epsilon_{k}+\mu+i\tilde{\Gamma}(-\omega)\right)\tilde{\xi}_{\kk}^\dagger(\omega)-\mu\tilde{\xi}_{-\kk}(-\omega)\\
-\mu\tilde{\xi}^\dagger_{\kk}(\omega)+\left(-\omega+\epsilon_{k}+\mu-i\tilde{\Gamma}^{*}(\omega)\right)\tilde{\xi}_{-\kk}(-\omega)
\end{array}\right).
\end{multline}
We get after tracing over the various baths the expression for the correlation matrix:
\begin{equation}
\label{eq_app:correlation tensor}
\mathcal{C}_{\kk}(\omega)=
\frac{1}{N_\kk(\omega)N_{-\kk}(-\omega)}\underbrace{\left(\begin{array}{cc}
M^{(11)}_\kk(\omega) & M^{(12)}_\kk(\omega)\\
M^{(21)}_\kk(\omega) & M^{(22)}_\kk(\omega)
\end{array} \right)}_{\equiv\mathcal{M(\omega)}},
\end{equation}
where
\begin{small}
\begin{subequations}
\begin{eqnarray}
N_\kk(\omega)&=&\left[\omega-\left(\epsilon_{k}+\mu+i\tilde{\Gamma}(\omega)\right)\right]\\
&&\qquad\qquad\times\left[\omega+\epsilon_{k}+\mu-i\tilde{\Gamma}^{*}(-\omega)\right]+\mu^2\nonumber\\
M^{(11)}_\kk(\omega)&=&S_{\rm{l}}(\omega_{BEC}+\omega)\left|\omega+\epsilon_\kk+\mu+i\tilde{\Gamma}(-\omega)\right|^2\\
&&\qquad\qquad+S_{\rm{p}}(\omega_{BEC}-\omega) \mu^2 ,\nonumber\\
M^{(21)}_\kk(\omega)&=&-S_{\rm{l}}(\omega_{BEC}+\omega)\\
&&\qquad\qquad\times\left[\omega+\epsilon_\kk+\mu+i\tilde{\Gamma}(-\omega)\right]\mu\nonumber\\
&&+S_{\rm{p}}(\omega_{BEC}-\omega)\left[\omega-\left(\epsilon_\kk+\mu+i\tilde{\Gamma}(\omega)\right)\right]\mu,\nonumber\\
M^{(12)}_\kk(\omega)&=&-S_{\rm{l}}(\omega_{BEC}+\omega)\\
&&\qquad\qquad\times\left[\omega+\epsilon_\kk+\mu-i\tilde{\Gamma}^{*}(-\omega)\right]\mu\nonumber\\
&&+S_{\rm{p}}(\omega_{BEC}-\omega)\left[\omega-\left(\epsilon_\kk+\mu-i\tilde{\Gamma}^{*}(\omega)\right)\right]\mu,\nonumber\\
M^{(22)}_\kk(\omega)&=&S_{\rm{l}}(\omega_{BEC}+\omega)\mu^2+S_{\rm{p}}(\omega_{BEC}-\omega)\\
&&\qquad\qquad\times\left|\omega-\left(\epsilon_\kk+\mu+i\tilde{\Gamma}(\omega)\right)\right|^2.\nonumber
\end{eqnarray}
\end{subequations}
\end{small}
To test the FDT it is also useful to calculate the ratios $\frac{\langle \hat{\Lambda}_\kk(\omega)\hat{\Lambda}^\dagger_{\kk}\rangle}{\langle\hat{\Lambda}^\dagger_\kk(\omega)\hat{\Lambda}_\kk\rangle}$ and $\frac{\langle \hat{\Lambda}_\kk(\omega)\hat{\Lambda}_{-\kk}\rangle}{\langle\hat{\Lambda}_{\kk}(-\omega)\hat{\Lambda}_{-\kk}\rangle}$. We obtain the following expressions:
\begin{align}
\frac{\langle \hat{\Lambda}_\kk(\omega)\hat{\Lambda}^\dagger_{\kk}\rangle}{\langle\hat{\Lambda}^\dagger_\kk(\omega)\hat{\Lambda}_\kk\rangle} & =
\frac{S_l(\omega_\text{BEC}+\omega)+S_p(\omega_\text{BEC}-\omega)A_k(\omega)}{S_p(\omega_\text{BEC}+\omega)+S_l(\omega_\text{BEC}-\omega)A_k(\omega)},\\
\frac{\langle \hat{\Lambda}_\kk(\omega)\hat{\Lambda}_{-\kk}\rangle}{\langle\hat{\Lambda}_{\kk}(-\omega)\hat{\Lambda}_{-\kk}\rangle} & =
\frac{S_l(\omega_\text{BEC}+\omega)+S_p(\omega_\text{BEC}-\omega)B_k(\omega)}{S_p(\omega_\text{BEC}+\omega)+S_l(\omega_\text{BEC}-\omega)B_k(\omega)},
\end{align}
with 
\begin{align}
A_k(\omega) &=\frac{\mu^2}{\left|\omega+\epsilon_\kk+\mu+i\tilde{\Gamma}(-\omega)\right|^2}, \\
B_k(\omega)&=\frac{-\omega+\epsilon_\kk+\mu-i\tilde{\Gamma}^{*}(\omega)}{\omega+\epsilon_\kk+\mu-i\tilde{\Gamma}^{*}(-\omega)}.
\end{align}
\section{Static correlations at low energy}\label{app:static-correlations}
 
In this Appendix, we calculate the static correlations at steady state in the low-energy regime $E_k\ll\Delta_{\text{diss}}$. In this regime, using the definition Eq.~(\ref{eq:z}) as well as the fact that $S_{\rm{l}}(\omega_{BEC})=S_{\rm{p}}(\omega_{BEC})$, we can approximate the expression Eq.~(\ref{eq_app:correlation tensor}) of the correlation matrix calculated in the previous Appendix as:
\begin{small}
\begin{subequations}
\label{eq_app:low-frequency-correlations}
\begin{eqnarray}
N_{\kk}(\omega)\simeq &=&\frac{1}{|z|^2}\left\lbrace\left[\omega-z\left(\epsilon_{k}+\mu\right)\right] \left[\omega+z^{*}\left(\epsilon_{k}+\mu\right)\right]+|z|^2\mu^2\right\rbrace\nonumber\\
&=&\frac{1}{|z|^2}(\omega-\omega_{\kk}^{+})(\omega-\omega_{\kk}^{-}),\\
M^{(11)}_\kk(\omega)&\simeq&\frac{S_{\rm{l}}(\omega_{BEC})}{|z|^2}\left[\left|\omega+z\left(\epsilon_\kk+\mu\right)\right|^2+|z|^2 \mu^2\right] ,\\
M^{(21)}_\kk(\omega)&\simeq& -2S_{\rm{l}}(\omega_{BEC})(\epsilon_{k}+\mu)\mu,\\
M^{(12)}_\kk(\omega)&\simeq&-2S_{\rm{l}}(\omega_{BEC})(\epsilon_{k}+\mu)\mu,\\
M^{(22)}_\kk(\omega)&\simeq&\frac{S_{\rm{l}}(\omega_{BEC})}{|z|^2}\left[\left|\omega-z\left(\epsilon_\kk+\mu\right)\right|^2+|z|^2 \mu^2\right],
\end{eqnarray}
\end{subequations}
\end{small}
where $\omega_{\kk}^{\pm}$ are the complex low energy mode frequencies of the condensate given by Eq.~(\ref{eq:mode-frequencies}). From these expressions, we can calculate  the dynamic structure factor $\mathcal{S}_{\kk}(t)$, which is defined as

\begin{equation}
\mathcal{S}_\kk(t)=\left(\begin{array}{cc}
\langle\hat{\Lambda}_\kk(t)\hat{\Lambda}^\dagger_{\kk}(0)\rangle & \langle \hat{\Lambda}_\kk(t)\hat{\Lambda}_{-\kk}(0)\rangle \\
\langle\hat{\Lambda}^{\dagger}_{-\kk}(t)\hat{\Lambda}^{\dagger}_{\kk}(0)\rangle & \langle\hat{\Lambda}^{\dagger}_{-\kk}(t)\hat{\Lambda}_{-\kk}(0)\rangle
\end{array}\right),
\end{equation}
and is related to the correlation matrix $\mathcal{C}_{\kk}(\omega)$ as $ \int_t \mathcal{S}_\kk(t) e^{-i\omega t}= \mathcal{C}_{\kk}(\omega)$.
Using a pole integration in the complex plane we obtain
\begin{multline}
\mathcal{S}_{\kk}(t)=\frac{-i|z|^2}{2(\omega_{\kk}^{+}-\omega_{\kk}^{-})(\omega_{\kk}^{+}+\omega_{\kk}^{-})}
\\
\left[\frac{\mathcal{M}(\omega_{\kk}^{+})\ee^{-i\omega_{\kk}^{+} t}}{2\omega_{\kk}^{+}}-\frac{\mathcal{M}(\omega_{\kk}^{-})\ee^{-i\omega_{\kk}^{-} t}}{\omega_{\kk}^{-}}\right],
\end{multline}
where $\mathcal{M}(\omega)$ has been defined in Eq.~(\ref{eq_app:correlation tensor}).
Setting $t=0$ we find the static correlation matrix :
\begin{equation}
\mathcal{S}_{\kk}(0)=\frac{-i|z|^2}{2(\omega_{\kk}^{+}-\omega_{\kk}^{-})(\omega_{\kk}^{+}+\omega_{\kk}^{-})}\left[\frac{\mathcal{M}(\omega_{\kk}^{+})}{2\omega_{\kk}^{+}}-\frac{\mathcal{M}(\omega_{\kk}^{-})}{\omega_{\kk}^{-}}\right].
\end{equation}
By injecting the expressions given by Eqs.~(\ref{eq_app:low-frequency-correlations}) as well as the explicit expressions for the condensate frequencies Eq.~(\ref{eq:mode-frequencies}), we find:
\begin{equation}
\mathcal{S}_{\kk}(0)=\frac{S_{\rm{l}}(\omega_{BEC})|z|^2}{2 z_{I}E_{k}^2}\left(\begin{array}{cc}
 \epsilon_k +\mu & -\mu  \\
 -\mu & \epsilon_k +\mu 
\end{array} \right).
\end{equation}
From Eqs.~(\ref{eq:z}),~(\ref{eq:effective-temperature}), we have that $\frac{z_{I}}{|z|^2} = \text{Im} \left(z^{-1}\right)=-\left.\frac{\dd \text{Re} (\widetilde{\Gamma}(\omega))}{\dd \omega}\right|_{\omega=0}=\frac{\beta_{\rm{eff}}\mathcal{S}_{\rm{l}}(\omega_{\text{BEC}})}{2}$, from which we deduce the final expression:
\begin{equation}
\mathcal{S}_{\kk}(0)=\frac{T_{\text{eff}}}{E_{k}^2}\left(\begin{array}{cc}
 \epsilon_k +\mu & -\mu  \\
 -\mu & \epsilon_k +\mu 
\end{array} \right).
\end{equation}

\end{document}